\documentclass[a4paper,fleqn,usenatbib]{mnras}

\usepackage{newtxtext,newtxmath}

\usepackage[T1]{fontenc}
\usepackage{ae,aecompl}

\usepackage{graphicx}	
\usepackage{amsmath}	
\usepackage{amssymb}	

\title[Tracing the accretion history of supermassive black holes through X-ray variability]{Tracing the accretion history of supermassive Black Holes through X-ray variability: results from the \textit{Chandra} Deep Field-South}

\author[M. Paolillo et al.]{M. Paolillo$^{1,2,3}$\thanks{E-mail: paolillo@na.infn.it},
          I. Papadakis$^{4,5}$,
          W. N. Brandt$^{6,7}$,
	  B. Luo$^{8}$,	  
	  Y. Q. Xue$^{9}$,
	  P. Tozzi$^{10}$,\newauthor
	  O. Shemmer$^{11}$,
	  V. Allevato$^{12,13}$,
	  F. E. Bauer$^{14,19,20}$,
	  A. Comastri$^{15}$,
	  R. Gilli$^{15}$,\newauthor
	  A. M. Koekemoer$^{16}$,
	  T. Liu$^{9,18}$,
	  C. Vignali$^{17,15}$,
	  F. Vito$^{6,7}$,
	  G. Yang$^{6,7}$,
	  J. X. Wang$^{9},$\newauthor
	  and X. C. Zheng$^{9}$\\
~\\
$^{1}$Dipartimento di Fisica ``Ettore Pancini'', Universit\`a di Napoli Federico II,via Cintia, 80126, Italy\\
         $^{2}$INFN - Unit\`a di Napoli, via Cintia 9, 80126, Napoli, Italy\\
         $^{3}$Agenzia Spaziale Italiana--Science Data Center, Via del Politecnico snc, 00133, Roma, Italy\\
         $^{4}$Physics Department, University of Crete, 710 03 Heraklion, Crete, Greece\\
         $^{5}$Foundation for Research and Technology - Hellas, IESL, Voutes, 71110 Heraklion, Crete, Greece\\
         $^{6}$Department of Astronomy \& Astrophysics, 525 Davey Lab, The Pennsylvania State University, University Park, PA 16802, USA\\
         $^{7}$Institute for Gravitation and the Cosmos, The Pennsylvania State University, University Park, PA 16802, USA\\
         $^{8}$School of Astronomy and Space Science, Nanjing University, Nanjing 210093, China\\
         $^{9}$CAS Key Laboratory for Research in Galaxies and Cosmology, Department of Astronomy, \\ University of Science and Technology of China, Hefei, Anhui 230026, China\\
          $^{10}$INAF -- Osservatorio Astrofisico di Firenze, Largo Enrico Fermi 5, I-50125 Firenze, Italy\\
          $^{11}$Department of Physics, University of North Texas, Denton, TX 76203\\
          $^{12}$Department of Physics, University of Helsinki, Gustaf H\"allstr\"omin katu 2a, FI-00014 Helsinki, Finland\\
          $^{13}$University of Maryland, Baltimore County, 1000 Hilltop Circle, Baltimore, MD 21250, USA\\
          $^{14}$Instituto de Astrof{\'{\i}}sica and Centro de Astroingenier{\'{\i}}a, Facultad de F{\'{i}}sica, Pontificia Universidad Cat{\'{o}}lica de Chile, Casilla 306, Santiago 22, Chile\\ 
	 $^{15}$INAF -- Osservatorio Astronomico di Bologna, via Gobetti 93/3, 40129 Bologna, Italy\\
          $^{16}$Space Telescope Science Institute 3700 San Martin Drive, Baltimore MD 21218, USA\\
          $^{17}$Dipartimento di Fisica e Astronomia, Universit\`a degli Studi di Bologna, via Gobetti 93/2, 40129 Bologna, Italy\\
          $^{18}$Astronomy Department, University of Massachusetts, Amherst, MA 01003, USA\\
	  $^{19}$Millennium Institute of Astrophysics (MAS), Casilla 306, Santiago, Chile\\
	  $^{20}$Space Science Institute, 4750 Walnut Street, Suite 205, Boulder, Colorado 80301 
}

\date{Accepted 11 July 2017. Received 10 January 2017}

\pubyear{2017}

\begin{document}
\label{firstpage}
\pagerange{\pageref{firstpage}--\pageref{lastpage}}
\maketitle

\begin{abstract}
We study the X-ray variability properties of distant AGNs in the \textit{Chandra} Deep Field-South region over 17 years, up to $z\sim 4$,  and compare them with those predicted by models based on local samples.  We use the results of Monte Carlo simulations to account for the biases introduced by the discontinuous sampling and the low-count regime. We confirm that variability is an ubiquitous property of AGNs, with no clear dependence on the density of the environment.
The variability properties of high-z AGNs, over different temporal timescales, are most consistent with a Power Spectral Density (PSD) described by a broken (or bending) power-law, similar to nearby AGNs. 
We confirm the presence of an anti-correlation between luminosity and variability, resulting from the dependence of variability on BH mass and accretion rate. We explore different models, finding that our acceptable solutions predict that BH mass influences the value of the PSD break frequency, while the Eddington ratio $\lambda_{Edd}$ affects the PSD break frequency and, possibly, the PSD amplitude as well. 
We derive the evolution of the average $\lambda_{Edd}$ as a function of redshift, finding results in agreement with measurements based on different estimators. The large statistical uncertainties make our results consistent with a constant Eddington ratio, although one of our models suggest a possible increase of $\lambda_{Edd}$ with lookback time up to $z\sim 2-3$. We conclude that variability is a viable mean to trace the accretion history of supermassive BHs, whose usefulness will increase with future, wide-field/large effective area X-ray missions.
\end{abstract}




\section{Introduction}
\label{Intro}
Flux variability is a defining characteristic of Active Galactic Nuclei (AGNs),
reflecting the small spatial region in which the observed emission is produced and the production mechanism itself \citep{Fabian79,Rees84}.
AGNs are observed to vary on all timescales, and across the whole electromagnetic spectrum, 
although the maximum power and fastest variations are found at the highest energies (X-rays and $\gamma$-rays), due to
the fact that such radiation is mainly generated close to the central engine and over small spatial regions \citep{Ulrich97}.

In the X-ray band in particular, extended and detailed observations of nearby sources, have revealed that the variability is characterised by `red noise' behaviour, with more power existing on the longest timescales, in close resemblance to what is observed in binary accreting systems containing smaller, stellar-mass Black Holes (BH).   
The origin of the X-ray variability itself is not well understood, and both internal (instabilities of the accretion flow, flaring corona, orbiting hotspots) and external (variable obscuration, micro lensing) phenomena have been proposed to explain the flux variations.
Early investigations of AGN variability did not show any distinct features in their Power Spectral Density 
 (PSD), suggesting that the PSD has a pure power-law shape \citep{Green93, Lawrence93} which unfortunately, has little power to discriminate among variability models. However several factors (analogies with galactic binaries, dependence of the emission 
on the physics of the accretion process, unphysical behaviour when extrapolated to long timescales) 
indicated that some characteristic timescale should be observable in the PSD.

More recently, the combination of long observing campaigns over several decades, and shorter high-quality \textit{XMM-Newton} observations, has allowed the discovery of at least one, and in some cases two, breaks in the PSD of nearby AGNs \citep{Uttley02, Papadakis02, Markowitz03, McHardy07}. Such features seem linked to both the BH mass and the properties of the accretion flow, and have enabled using variability to test the properties of the accretion flows as well as to measure the main physical parameters (BH mass, accretion rate) of the AGN. Similarly, we have been able to see extreme cases of variability induced by varying column densities of obscuring material, in type 1 \citep{Yang16}, type 2 AGNs \citep[e.g.][]{Nardini11,Risaliti11,Giustini11,Risaliti09} and in BAL quasars \citep[e.g.][]{Lundgren07,Gibson08,Gibson10}. 

Most of our knowledge about AGN variability in the X-ray band is derived from extensive observations of nearby and mostly low-luminosity 
AGNs, as these were the only ones initially accessible by low-effective area and/or low spatial resolution instruments. Such facilities have been the only ones allowing the long and regular monitoring campaigns required to avoid the problems introduced by low statistics and irregular sampling in the temporal analysis. The extension of such studies to a larger population
of distant sources requires both a large effective area and a good angular resolution to avoid crowding effects. 
Progress was made adopting a less sophisticated approach, measuring the integrated power over long timescales in an attempt to investigate the variability properties of AGNs over cosmological volumes. For instance \citet{Almaini,Manners02} studied samples of QSOs selected from \textit{ROSAT} surveys up to $z\lesssim 4$; \citet{Paolillo04} analysed the variability properties of AGNs in the \textit{Chandra} Deep Field-South (CDF-S) using \textit{Chandra} data, \citet{Papadakis08} and \citet{Allevato10} used \textit{XMM-Newton} observations to study the variability of AGNs in the Lockman Hole and the CDF-S respectively, \cite{Vagnetti11} and \citet{Middei17} investigated serendipitous \textit{XMM-Newton},\textit{Swift} and \textit{ROSAT} samples, while \citet[2017, submitted]{Shemmer14} explored a group of luminous quasars combining \textit{ROSAT}, \textit{Chandra} and \textit{Swift} observations. 

These works have shown that variability is ubiquitous in AGNs and that it has similar properties to those of nearby and less luminous AGNs, but also suggested that its amplitude may increase with lookback time and is possibly a tracer of the higher average accretion rates present in the earlier Universe.
The results so far are not conclusive and suffer from biases due to sparse sampling and low statistics, as well as randomness intrinsic to red-noise processes. For instance, \citet{Gibson12}, studying a serendipitous sample of SDSS spectroscopic quasars, confirmed several results of previous works but failed to detect any clear evidence of increased variability at large redshifts. Similar conclusions were reached by other authors: \citet{Mateos07} and \citet{Lanzuisi14} studying the \textit{XMM-Newton} light curves of AGNs in the Lockman Hole and COSMOS fields, and by \cite{Vagnetti16} using the MEXSAS serendipitous sample.

The CDF-S represents the deepest observation of the Universe in X-rays. As discussed above, the first 1 Ms data were used by \cite{Paolillo04} to investigate the nature of variability in distant AGNs, but many of their results were only marginally significant due to the low number of sources, the limited availability of spectroscopic redshifts and the limited timescale coverage. This dataset has grown over time to span, with the 7 Ms data presented in \citet{Luo16}, a time interval of $\sim 17$ years, reaching a depth of $1.9\times 10^{-17}$, $6.4\times 10^{-18}$ and $2.7\times 10^{-17}$ erg cm$^{-2}$ s$^{-1}$ for the $0.5-7$, $0.5-2$ and $2-7$ keV bands respectively, and has accumulated a wealth of ancillary multi-wavelength data. This work is thus intended to test and extend the previous results, and link them to our knowledge based on nearby samples. We have already exploited these data in part in \citet{Shemmer14} where we used the 2 Ms CDF-S light curves to compare the bulk variability of radio-quiet AGNs to bright high-redshift quasars,  in \citet{Young12} where we used the 4 Ms data to detect the faint AGN population in normal galaxies by means of variability, and in \citet{Yang16} to investigate the long-term variability of AGNs. Here we present a more refined analysis of the 7 Ms light curves probing different temporal timescales in order to understand the connection between variability and AGN physical properties at $z>0.5$. 

The paper is organised as follows: in \S \ref{sec_data} we discuss the data and the lightcurves extraction process, \S \ref{var_det} explains how we detect and characterise the population of variable sources, \S \ref{ex_var} explains how we measure the average variability and study its dependence on the  properties of the AGN population, \S \ref{accretion_section} discusses how we test different variability models and use them to constrain the AGN accretion history with lookback time. Finally, in \S \ref{disc_sec}, we discuss our results and present our main conclusions.

Throughout the paper we adopt values of $H_0 = 70$ km s$^{-1}$ Mpc$^{-1}$, $\Omega_M = 0.3$, and $\Omega_\Lambda = 0.7$ \citep{Spergel03}.
  

\section{The data}
\label{sec_data}
The CDF-S data used here are those described in detail in
\citet[][also see \citealt{Luo08, Xue11}]{Luo16}. For completeness we shortly summarize here the main properties of the dataset, referring the reader to the above papers for a thorough discussion of the data properties. The dataset consists of 102 observations collected by \textit{Chandra} between 1999 and 2016, adding up to a total exposure time of 6.727 Ms; the individual observations have exposure times ranging from $\sim 9$ ks up to $141$ ks, and have very similar aimpoints within $\sim 1'$, although different roll angles (see Table~1 in \citealt{Luo16}). The data reduction procedure adopted in order to create event lists and exposure maps is described in detail in \citet{Luo16} and we refer the reader to that paper for details.

\begin{figure*}
   \centering
   \includegraphics[bb=405 18 575 760,width=0.22\textwidth, angle=90]{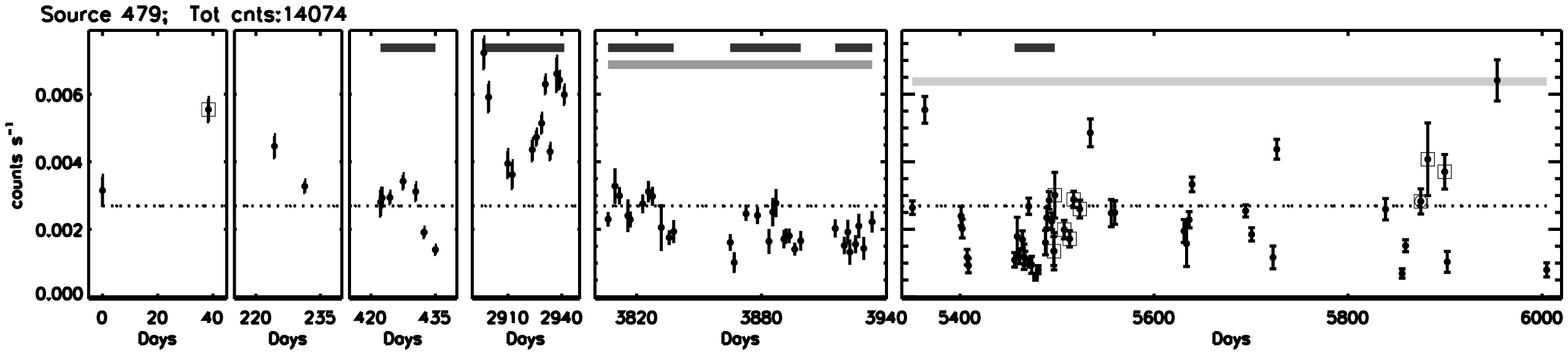}
   \includegraphics[bb=405 18 575 760,width=0.22\textwidth, angle=90]{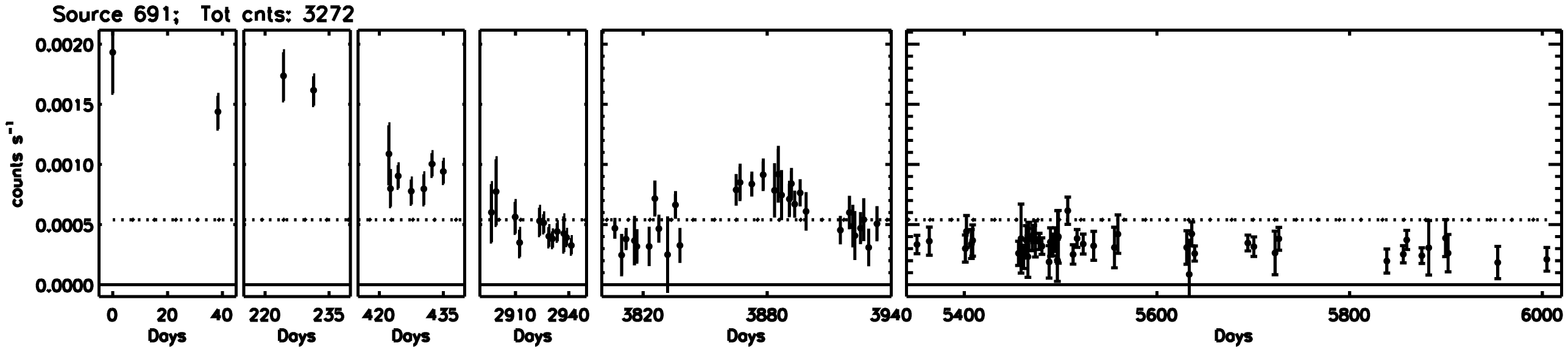}
   \includegraphics[bb=405 18 575 760,width=0.22\textwidth, angle=90]{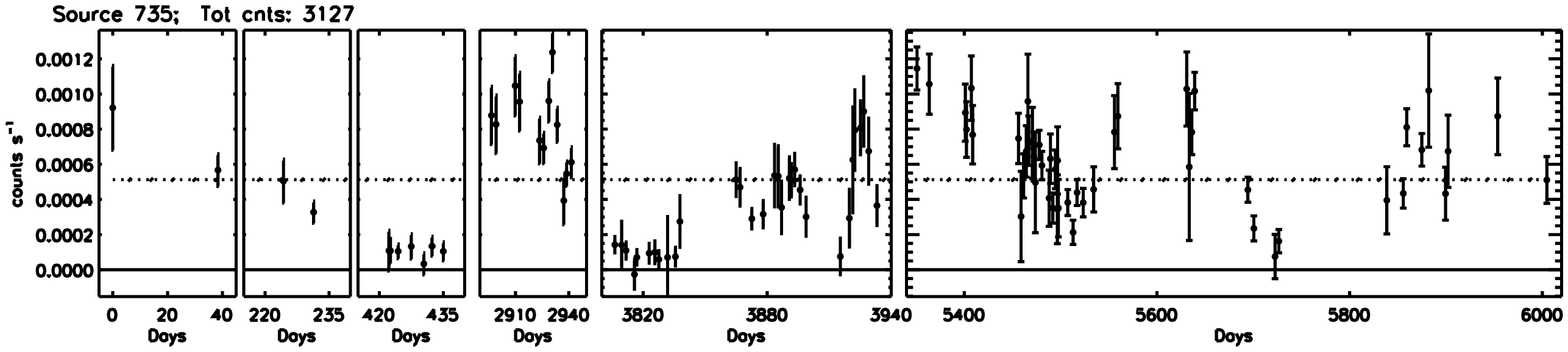}
   \includegraphics[bb=405 18 575 760,width=0.22\textwidth, angle=90]{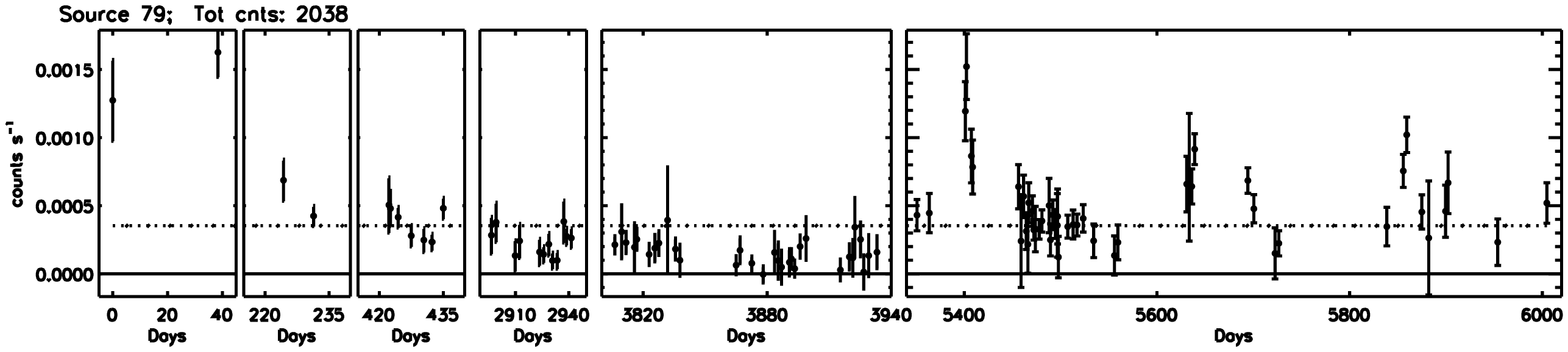}
   \includegraphics[bb=405 18 575 760,width=0.22\textwidth, angle=90]{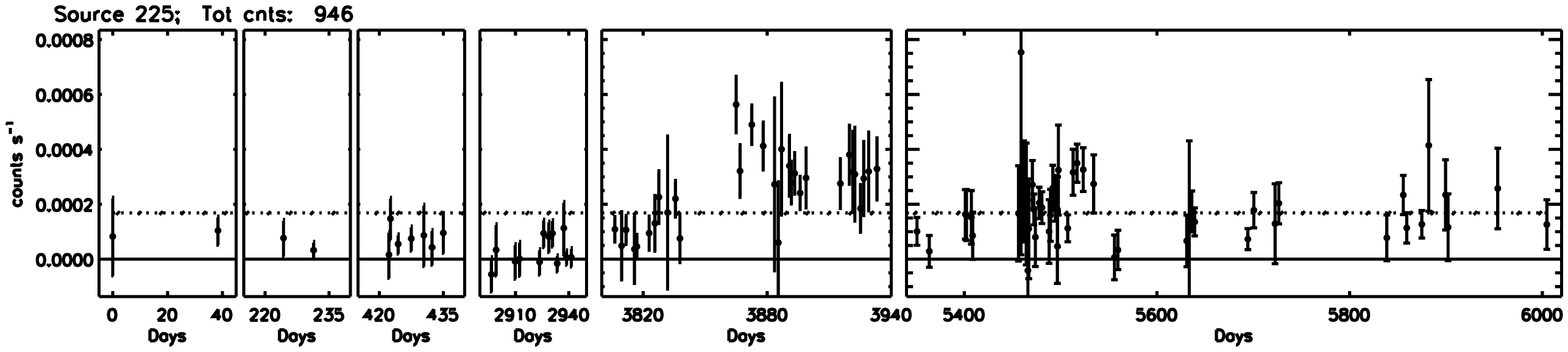}
      \caption{Example of CDF-S lightcurves in the 0.5-8 keV band, for sources with different total counts and temporal behaviour; given the sparse cadence of CDF-S observations, each panel groups nearby observations separated by large temporal gaps; specifically the first 3 panels represent the 1st Ms data presented by \citet{Giacconi02} and used in \citet{Paolillo04}, the 4th, 5th and 6th panels cover the additional 1, 2 and 3 Ms presented respectively by \citet{Luo08, Xue11,Luo16}. The average count rate is marked by a dotted line; a solid line shows the zero count rate level. The source number from the \citet{Luo16} catalog and the total source counts are shown in the upper left corner of each row. The squares, in source 367, mark observations where part of the source background falls outside the FOV. The dark, medium and light grey bars in the top row highlight the temporal segments used to compute variability on short, intermediate and long timescales, as discussed in \S \ref{modelingagn}. }
         \label{CDF-S_source_ltc}
\end{figure*}

In order to study the temporal behaviour of the AGN population, we extract AGN lightcurves following the same procedure adopted by \cite{Paolillo04} for the 1 Ms dataset. 
We start from the main source catalog of \cite{Luo16}, consisting of 1008 X-ray sources, which represent our ``main sample''; we ignore instead the Supplementary Near-Infrared Bright Catalog in \cite{Luo16}, as these sources are all too faint to be useful in our analysis. For each source we measured counts within a circular aperture with variable radius $R_S$ depending on the angular distance $\theta$ in arcsec, from the average aimpoint: $R_S=2.4\times FWHM$ arcsec, where $FWHM=\sum_{i=0,2} a_i\theta^i$ is the estimated full width half maximum of the point spread function (PSF) and $a_i=\{0.678,-0.0405,0.0535\}$ \cite[see][for further details]{Giacconi02}. The only difference with respect to \cite{Paolillo04} is that the minimum radius was set to 3 arcsec, in order to exploit fully the sharp \textit{Chandra} PSF in the FOV center, and minimise the cross contamination between nearby sources\footnote{Note that this was not an issue in the 1 Ms data where, due to the lower sensitivity, crowding was less severe.}. Similarly the local background for each source was measured in a circular annulus of inner (outer) radius $R_S+2$ ($R_S+12$) arcsec. Neighbouring objects were always removed from the source or background region when they overlapped. This approach is less sophisticated than the one adopted by \cite{Luo16} who used the ACIS EXTRACT software to model the \textit{Chandra} PSF and extract fluxes within polygonal regions; however it has the advantage of being simpler and avoiding the low S/N  wings of the \textit{Chandra} PSF \cite[also see][]{Vattakunnel12}. A comparison between our total fluxes and those derived in  \cite{Luo16} shows that on average our extraction procedure recovers 95\% of the total source flux. In any case we stress that we use our aperture photometry only for the variability analysis, while turning to the more accurate \cite{Luo16} photometry to obtain total fluxes and luminosities.

We binned the data into individual observations: this allows derivation of lightcurves with 102 points over an 17 years interval. Sources near the edge of the detector may be missing in several observations due to the different aim-point and roll-angles of each pointing. We follow \cite{Paolillo04} in retaining only the epochs where $>90\%$ of the source region and $>50\%$ of the background region falls within the FOV. In any case 884 (88\%) of our sources have lightcurves with at least 50 bins and 758 (75\%) are sampled by all 102 observations.  
The lightcurves were extracted both in the full 0.5-8 keV band, and in the 2-8 keV rest-frame band for the 986 sources with available redshifts\footnote{In the case of the 9 objects ($<1\%$ of the total sample) with $z>4$, more than 10\% of the rest-frame 2-8 keV band falls outside the observed band.}. Examples of CDF-S lightcurves are shown in Figure~\ref{CDF-S_source_ltc}. A short movie showing the variability of sources in the entire CDFS field can be found at http://people.na.infn.it/paolillo/MyWebSite/CDFS.html .

\section{Finding variable AGNs}
\label{var_det}
\begin{figure}
   \centering
   \includegraphics[bb=0 25 605 595, width=0.48\textwidth]{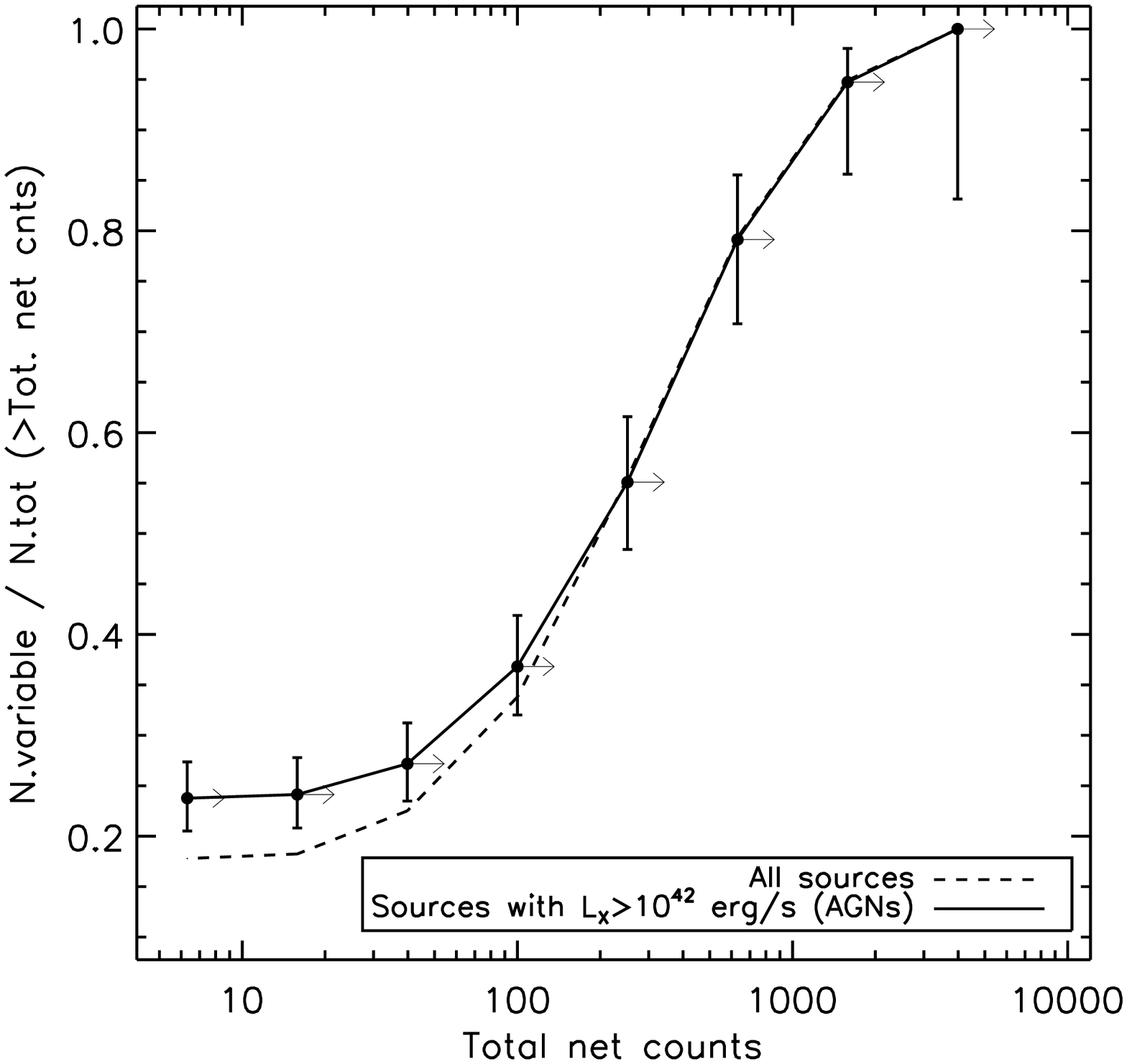}
      \caption{Cumulative fraction of variable sources in the CDF-S, as a function of the net source counts. The number of variable sources increases at higher counts (i.e., S/N) We show both the entire main sample, and the subsample of sources with $L_X>10^{42}$ erg s$^{-1}$ which minimises contamination by normal galaxies. Error bars represent the 95\% binomial uncertainty (calculated using the Bayesian approach of \citealt{Cameron11}).}
         \label{var_frac}
\end{figure}

To assess the significance of variability of the sources in the main sample, we compute the $\chi^2$ of each lightcurve defined as $$\chi^2=\frac{1}{N_{obs}-1}\sum_{i=1}^{N_{obs}}\frac{(x_i-\bar{x})^2}{\sigma_{err,i}^2}$$  
where $N_{obs}$ is the number of observations in which the sources fall inside the FOV, $x_i$ and $\sigma_{err,i}$ are the count rate and its error measured in the \textit{i}th observation after background subtraction, and correcting for exposure and effective area variations\footnote{http://cxc.harvard.edu/ciao/why/acisqecontam.html}, and $\bar{x}$ is the average count rate extracted from the stacked 7 Ms data. We then compare the measured $\chi^2$ with the expected value based on a set of 1000 simulations of each source, assuming a constant flux, as done in \cite{Paolillo04}. The simulations reproduce all the actual data properties, including Poisson noise, background and exposure. This allows accounting for the very large deviation from Gaussianity which affects the low-count regime, and prevents the use of any analytical expression based on such an assumption. We flag as variable, sources with $P(<\chi^2)>95\%$, finding that 165 out of 1008 (16\%) of the sources are variable. This fraction however is affected by the low statistics for the majority of the sources (70 median counts) and, to a lesser extent, the contamination at low fluxes by normal galaxies with $L_X\lesssim 10^{42}$ erg s$^{-1}$ (where $L_X$ is the rest-frame 0.5-8 keV luminosity calculated as described below).  
In Figure~\ref{var_frac} we plot the cumulative fraction of variable sources, showing that at high count levels all sources are found to vary. The plot confirms the trend observed by \cite{Paolillo04} in the 1 Ms dataset, and \cite{Young12} and \cite{Yang16} in the 4 and 6 Ms data with lower time resolution, that variability is more easily detected in higher S/N sources and supports the view that all AGNs are intrinsically variable on a broad range of timescales.

We note that \cite{Paolillo04} implemented also an additional variability estimate ($\chi^2_{max}$) based on the maximum deviation from the mean, observed among all the bins in the lightcurve. This approach was useful both because in the 1 Ms dataset we had at most 11 points in each lightcurve and thus low variability levels could be hard to detect, and also to detect short transient events. The 7 Ms dataset however has much better sampled lightcurves, and spans longer timescales, thus probing lower frequencies where the variability power of AGNs is expected to be larger due to the red-noise PSD. 
Moreover \citet{Luo08b} searched the first 2 Ms of CDF-S data, finding no evidence that short-duration events (durations of few months, such as stellar tidal disruptions) may dominate the observed variability, and only a few fast-transient has been observed over the full 7 Ms data \citep[Zheng et al., in preparation]{Bauer17}. In this work we thus concentrate on variability estimates which are averaged over many epochs.

\begin{figure}
   \centering
   \includegraphics[bb=0 20 720 595, clip, width=0.48\textwidth]{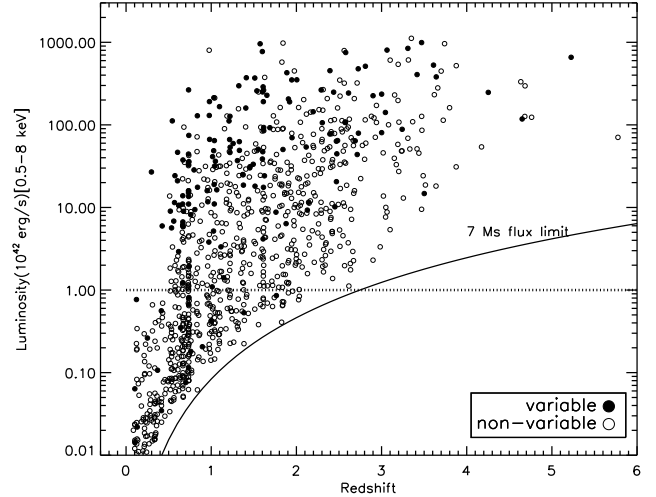}
      \caption{X-ray luminosity vs redshift. Solid dots mark variable sources. 
The horizontal dotted line marks the $L_X=10^{42}$ erg s$^{-1}$ limit generally used to discriminate AGNs from galaxies while the solid line shows the 7 Ms flux limit.}
         \label{Lx_z}
\end{figure}

The X-ray luminosity vs redshift distribution of sources from the 7 Ms \cite{Luo16} sample is presented in Figure~\ref{Lx_z}. The rest-frame 0.5-8 keV luminosity was calculated by \citet{Luo16}  modelling the X-ray emission using a power-law with both intrinsic and Galactic absorption;  the  column density was constrained finding the value that best reproduced the observed hard-to-soft band ratio, assuming an intrinsic power-law photon index of $\Gamma_{int}=1.8$ for AGN spectra. 
Variable sources (solid circles) are detected up to $z \sim 5$, but lie preferentially among the brightest sources at any redshift due to the large number of counts required to detect flux changes (see also Figure~\ref{exv_cnts}).
As was the case also in the 1 Ms data, there are several variable sources below the $L_X=10^{42}$ erg s$^{-1}$ limit, often adopted to separate AGNs from normal galaxies; this is not surprising since many galaxies are expected to host low-luminosity AGNs whose emission significantly contributes to the overall galaxy X-ray luminosity. In fact \cite{Young12} already searched the 4 Ms data to identify LLAGNs, using longer integration timescales ($4\times 1$ Ms bins) in order to increase the likelihood of detecting variability in faint sources.

%

\section{Measuring the AGN variability amplitude}
\label{ex_var}

To quantify the variability power we compute the normalized excess variance, as defined by
\citep{Nandra97,Turner99}:
\begin{equation}
 \sigma^2_{NXS} = \frac{1}{N \overline{x}^2} \sum_{i=1}^{N} \,
                  [(x_i - \overline{x})^2 - \sigma_{err,i}^2],
 \label{eq:exvar2}
\end{equation}
where $x_i$ and $\sigma_{err,i}$ are, again, the (exposure time and effective area corrected) count rate and its error in i-th bin, $\overline{x}$ is the average count rate of the source from the stacked 7 Ms data, and $N$ is the number of bins used to estimate $\sigma^2_{NXS}$.
\cite{Almaini} note that the excess variance, as defined above, is a maximum likelihood (ML) estimator of the intrinsic lightcurve variance only in the case of identical normally distributed errors; if this is not the case the authors point out that there is no exact analytic ML solution that allows estimation of the intrinsic variance thus requiring a numerical approach. However, \cite{Allevato13} have shown that in practical applications, with realistic lightcurves and sparse sampling, the two approaches yield identical results, as expected by the fact that the  sources of uncertainties described below are much larger than those introduced by the use of an approximate solution. For such reason we prefer to use the excess variance as commonly done in the literature. Furthermore, while \citet{Antonucci14} and \citet{Vagnetti16} warn about possible biases introduced by the comparison of the excess variance in sources at different redshifts, we note that this bias only originates from an improper use of this estimator if one does not account for the different rest-frame timescales.

The formal error on $\sigma^2_{NXS}$ is given by \citet{Turner99}, assuming stationarity and uncorrelated Gaussian processes. 
Subsequently \citet{Vaughan03} provided an alternative approach, more suited to compare the temporal behaviour in different energy bands. These estimates however only account for measurement errors and not for the random scatter intrinsic to any red-noise process. In addition, as shown in \cite{Allevato13}, the irregular sampling pattern in the case of sparsely sampled lightcurves should also introduce additional scatter that will depend on the sampling scheme and the intrinsic (and a-priori unknown) PSD shape.

\begin{figure}
  \centering
  \includegraphics[bb=40 1 613 613,clip,width=0.5\textwidth]{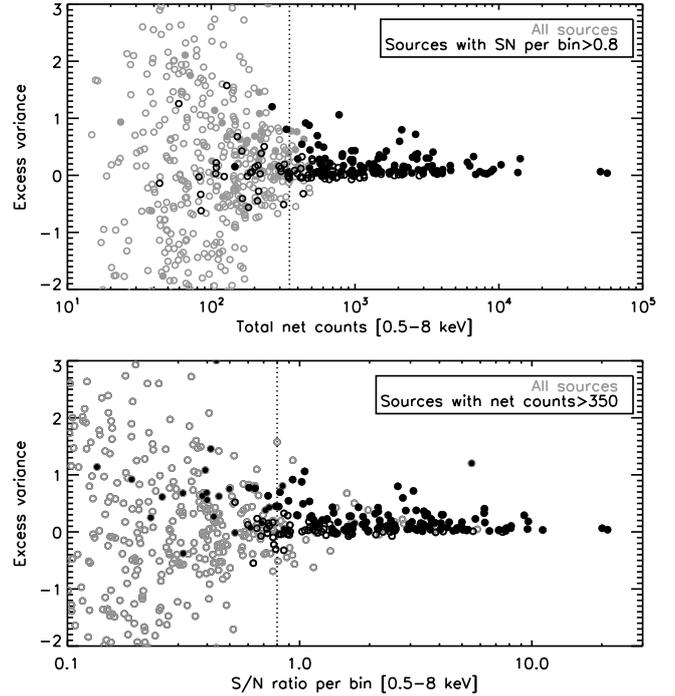}
     \caption{Excess variance vs total source counts (top panel) and S/N ratio per bin (bottom panel) in the 0.5-8 keV band. Variable and non-variable sources are plotted as solid and open circles respectively. Note that a some (faint) sources lie outside the plotted range. The vertical dotted lines show the counts and S/N limits adopted in this work to ensure that the observed variability is not dominated by statistical uncertainties.}
        \label{exv_cnts}
\end{figure}

Figure~\ref{exv_cnts} shows the excess variance of the main sample as a function of total source counts and average S/N ratio per bin\footnote{This is simply the S/N of the source computed for each time bin (i.e. observation) and then averaged over all time bins composing the full lightcurve. This quantity is a more robust measure of the quality of the data on the shortest timescales and is the same used in the simulations of \citet{Allevato13}.} in the full \textit{Chandra} energy band. At low count rates or low S/N ratios there is a very large scatter in the excess variance and a significant fraction of sources have negative values. As the S/N increases  (average S/N per bin $\gtrsim 0.8$, corresponding roughly to total counts$\gtrsim 350$, see Figure~\ref{exv_cnts}) the distribution skews significantly toward positive excess variances, reflecting the improved ability to measure the intrinsic source variance. We measure a median variance $\sigma^2_{NXS}=0.14_{-0.08}^{+0.16}$ corresponding to count rate fluctuations of $\sim 40\%$ ($\sigma_{NXS}=0.37$) for variable sources with $>350$ counts, where the uncertainties are the lower and upper quartiles. This is $\sim 30\%$ larger than observed by \citet[where $\sigma_{NXS}=0.28$]{Paolillo04}, as expected if AGNs have a red-noise PSD whose power increases on the longer timescales sampled here, and possibly also due to the fact that we are probing fainter sources\footnote{In \cite{Paolillo04} we adopted a threshold of 100 counts for 1 Ms observing time.} which tend to be intrinsically more variable (see \S \ref{L_var_corr}).

Motivated by the discussion in the previous paragraph, we decided to create two new samples of sources with: 1) S/N per bin $>0.8$, in the observed (0.5--8 keV) and the rest-frame (2--8 keV) bands, and 2) more than 90 points in their lightcurves (to exclude sources at the edge of the field-of-view sampled only by part of the observations). We name them as the the "bright-O" and "bright-R" samples, respectively. The S/N lower limit value of 0.8 is reinforced by the results of \citet{Allevato13} who showed that such a threshold is necessary to measure accurately the excess variance in sparsely sample data, as long as we average 10-20 individual measurements. For the bright-R sample all quantities, including luminosities and excess variance, are computed in the rest-frame 2--8 keV band. 

We also used the Extended \textit{Chandra} Deep Field-South radio catalog by \citet{Bonzini13}, which classifies radio sources based on their infrared 24 $\mu$m to radio 1.4 GHz flux density ratio, to identify radio-loud AGNs in our samples (see their \S 3.2). There are 6 and 5 radio-loud AGN in the bright-O and bright-R samples, respectively. Although we do not detect a significant difference in the average variability of such sources from the rest of the AGN population, we decided to remove them anyway from the subsequent analysis as the physical origin of the variability may be different (e.g. originating from the jet). The final number of sources in the bright-O and bright-R samples is 110 and 94, respectively. The different sizes of the two samples is due to the presence of sources with missing redshift and to the lower average S/N in the rest-frame band.

\begin{figure*}
  \centering
    \includegraphics[width=0.5\textwidth]{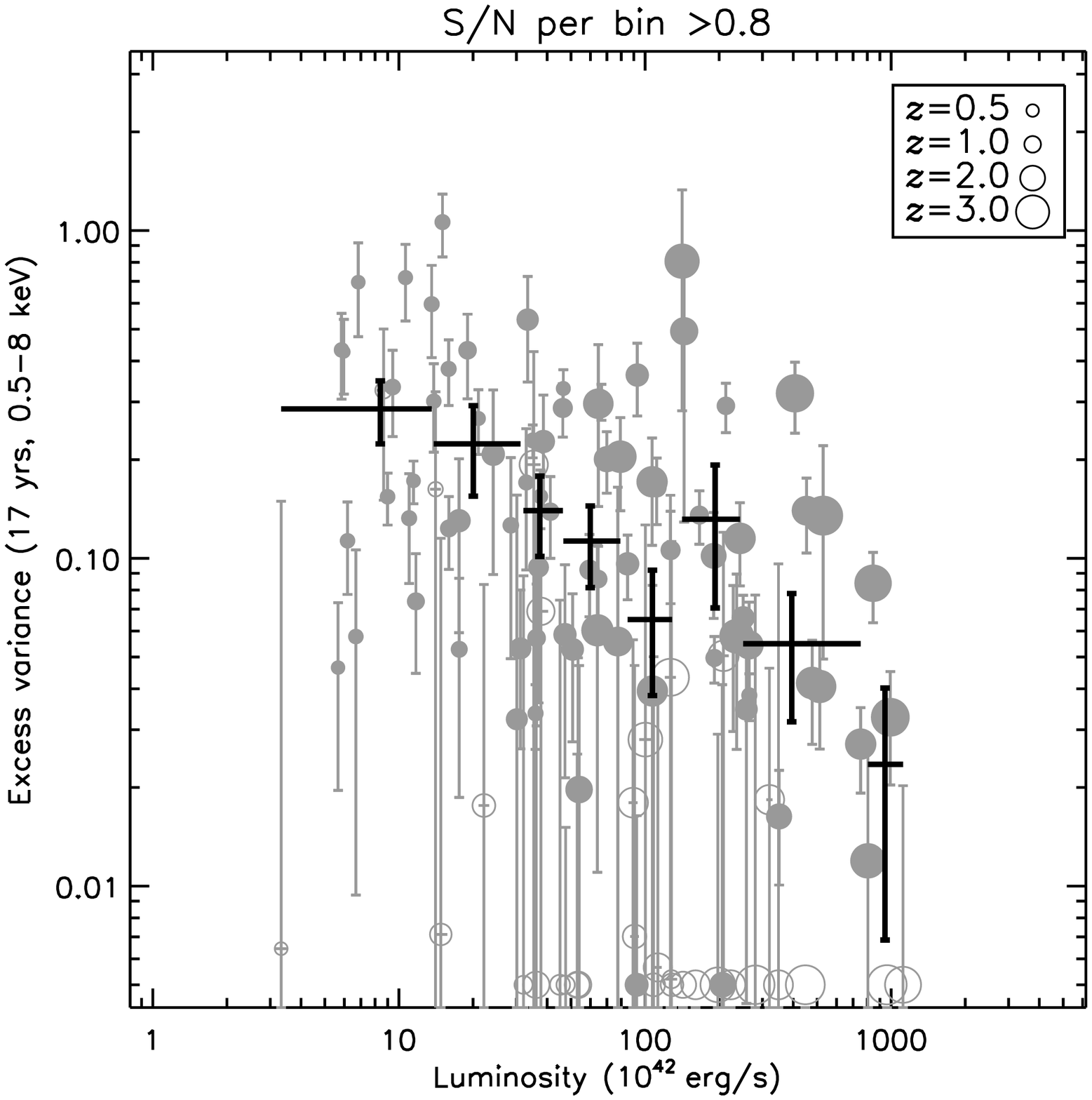}\includegraphics[width=0.5\textwidth]{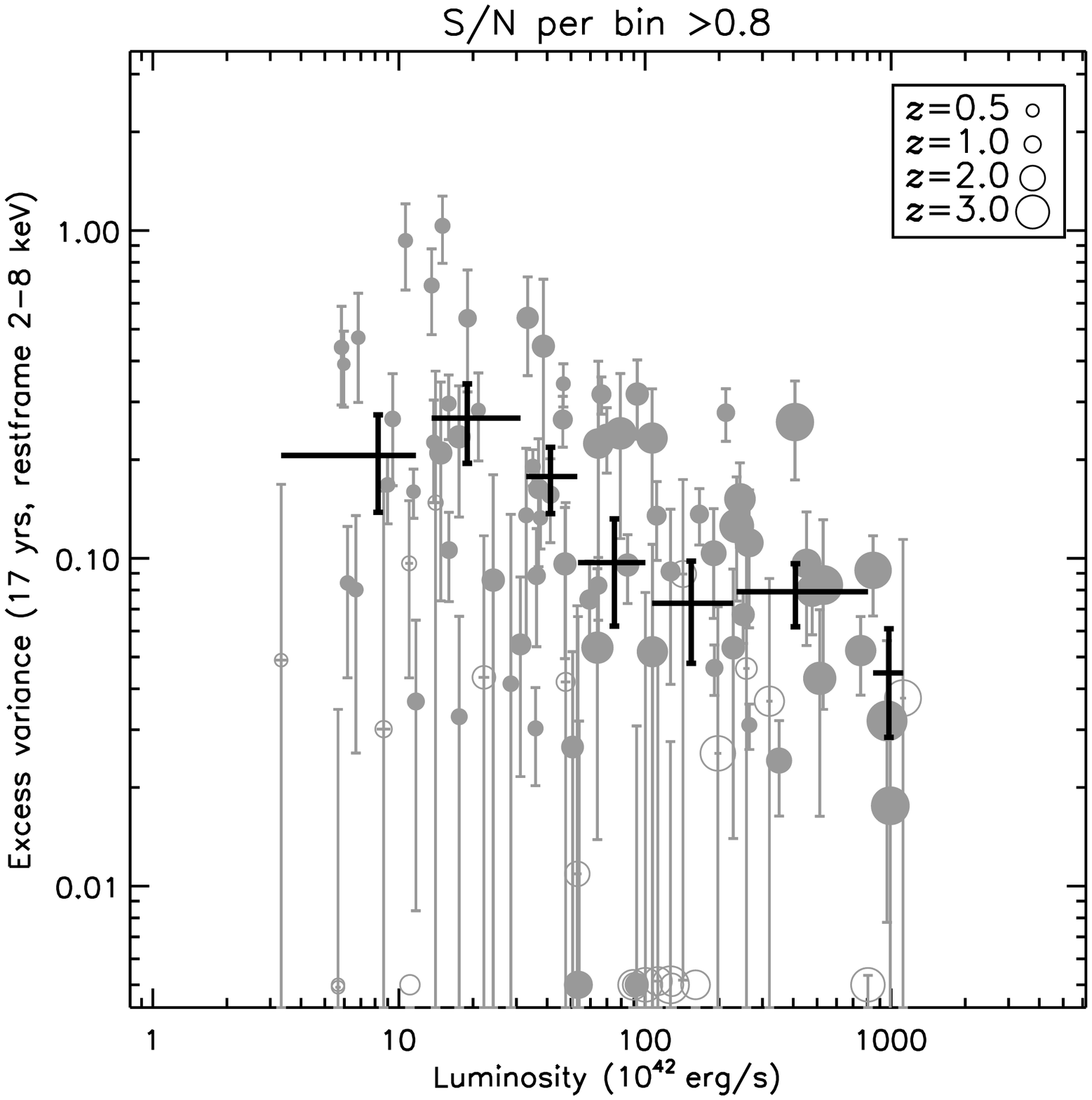}
    \caption{Excess variance over the full 7 Ms (17 years span) vs luminosity for the bright-O and bright-R samples (left and right panel, respectively). The individual sources are plotted as light grey circles of increasing size as a function of redshift. Black crosses show the average variance and its error in bins with $\sim 15$ sources. Variable/non-variable sources are shown as filled/empty symbols, respectively; sources with excess variance $\leq 5\times 10^{-3}$ are shown at the $5\times 10^{-3}$ level. The S/N threshold reported in the Figure titles refers to the bins in the light curves, as explained in the text. }
        \label{Lx_var}
\end{figure*}

\subsection{Dependence of AGN variability on luminosity, redshift and local density}
\label{L_var_corr}
In Figure~\ref{Lx_var} we plot the measured $\sigma_{NXS}^2$ against the (absorption-corrected) X-ray luminosity for sources in the bright-O and bright-R sample (grey points). The black crosses show the mean $\sigma^2_{NXS}$ (and luminosity) over 15 sources. This mean should be representative of the intrinsic excess variance (assuming that all sources in a given luminosity bin have a similar intrinsic $\sigma^2_{NXS}$). We point out that it is a common mistake to remove non-variable or negative $\sigma^2_{NXS}$ sources. As shown in \citet{Allevato13}, the $\sigma^2_{NXS}$ distribution extends to negative values especially for low variability/low S/N sources; removing these values would bias the variability ensemble estimates. We therefore used the excess variance measurements of all sources  in each bin, irrespective of whether they are positive or negative, to estimate the mean $\sigma^2_{NXS}$. The error of the individual points in Fig.\,\ref{Lx_var} take into account the measurement error of the points in the light curves, only, and have been estimated using the equations in \citet{Turner99}, as discussed above. Instead, the error of the mean excess variance in each bin are estimated following \citet{Allevato13}, and should be representative of the true, overall uncertainty of the mean excess variance.

Figure \ref{Lx_var} shows that AGN variability is anti-correlated with (unabsorbed) X-ray luminosity, thus confirming the results obtained in previous investigations of the CDF-S on different timescales by \cite{Paolillo04,Young12,Shemmer14,Yang16}.
The two panels in the Figure show that the dependence of the variability amplitude on X--ray luminosity is very similar between the bright-O and bright-R samples. However, in order to: 1) avoid the effects due to the absorbing column and its variations in time, which mainly influence energies below 2 keV, 2) to eliminate complications in the interpretation of our results due to the differences in the PSD between the soft and hard band seen in some local AGNs \citep[e.g.][]{McHardy04b}, and 3) to allow a better comparison of our results with those from with variability studies of local AGNs on long time scales, which are mainly based on {\it RXTE} data and focus on the 2--10 keV energy range, from now on, we will only use the rest-frame $2-8$ keV measurements of the bright-R sample.

\begin{figure}
  \centering
  \includegraphics[bb=0 35 730 600, clip, width=0.5\textwidth]{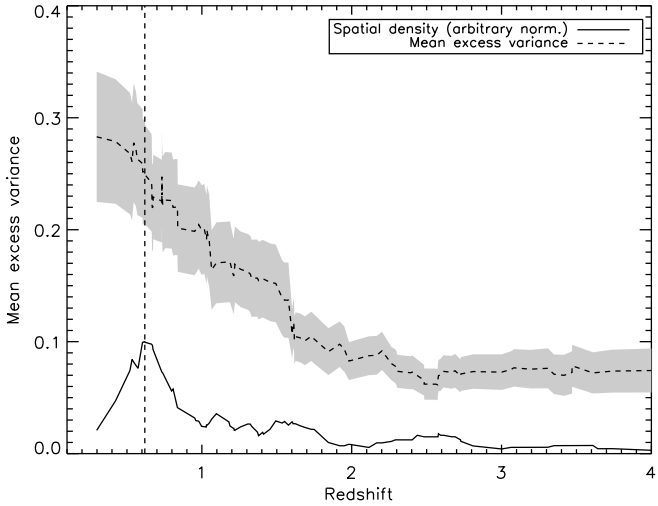}
     \caption{Running average excess variance (in bins of 20 sources) as a function of redshift; the error on the mean (68\% uncertainty) is shown by the grey shaded area. The solid line shows the volume density of sources, with an arbitrary normalization. The vertical dashed line indicates the position of the spatial density peak at $z\simeq 0.6-0.7$. }
        \label{varfrac_z_0.1}
\end{figure}

Figure~\ref{varfrac_z_0.1} shows the dependence of the average excess variance (including both variable and non-variable sources) on redshift. The dashed line in this figure shows the running average of the excess variance measurements of the sources in the bright-R sample and its error (estimated as explained above). We observe the variability amplitude to decrease with increasing redshift, which is consistent with the fact that at hight-redshift we probe higher luminosity sources (see Figure~\ref{Lx_z}) which are intrinsically less variable (Figure~\ref{Lx_var}).

It is interesting to compare the dashed and the solid lines in Fig.\,~\ref{varfrac_z_0.1}. The solid line indicates the volume density of the CDF-S sources as a function of redshift (in arbitrary units, so that it can be easily compared with the dashed line). The CDF-S region is characterised by several overdensities in redshift space, due to the presence of large-scale structures \citep[e.g.][]{Gilli03}. The most prominent one, at $z\simeq 0.7$, also contains a large number of variable sources down to $L_X\simeq 10^{41}$ erg s$^{-1}$. The comparison of the mean excess variance with the local volume density does not any correlation above the statistical uncertainty.
The lack of any excess of the average excess variance, coincident with the $z\simeq 0.7$ density peak, suggests that the variability amplitude in these AGN is not affected by environmental effects that could trigger and enhance variability through, e.g. enhanced accretion processes or dynamical instabilities.

\subsection{Variability dependence on timescale}
\label{var_timescale}

In addition to luminosity and redshift, if the intrinsic variability process has a ``red-noise" character, the excess variance also depends on the rest-frame duration of the light curves, which usually span a fixed time interval in the observer's frame. To investigate this issue, we computed the excess variance of each object in the bright-R sample on 4 different timescales (in the observer's frame): 6005 days, 654 days,128 days and 45 days (the `7 Ms', `long', `intermediate' and `short' timescales, respectively). The first time interval is simply the total duration of the full 7 Ms dataset. The `long' timescale measurement was obtained using data only from the last 3 Ms which correspond to the points covered by the light grey bar in the top row of Figure~\ref{CDF-S_source_ltc}, between 5350 and 6000 days. The `intermediate'  excess variance was measured over the interval 2-4 Ms, i.e. between 3800 and 3940 days (intermediate grey bar in Figure~\ref{CDF-S_source_ltc}). The excess variance of the shortest timescales is the hardest to estimate since the variations on these timescales are usually dominated by statistical noise. To increase the reliability of our measurement we averaged the variance measured from 6 different short time intervals (420-450, 2900-2950, 3800-3840, 3860-3900, 3910-3940, 5455-5500 days, dark grey bars in Figure~\ref{CDF-S_source_ltc}) where the \textit{Chandra} observations had a dense cadence and the sampling is more uniform. As shown by Vaughan et al. (2003) and Allevato et al. (2012) averaging over multiple observations allows to reduce the the intrinsic scatter on $\sigma^2_{NXS}$. To reduce further the uncertainty of the excess variance estimates, we limited the analysis to objects whose lightcurves have an average S/N per bin $>1.5$.

\begin{figure}
  \centering
  \includegraphics[width=0.48\textwidth]{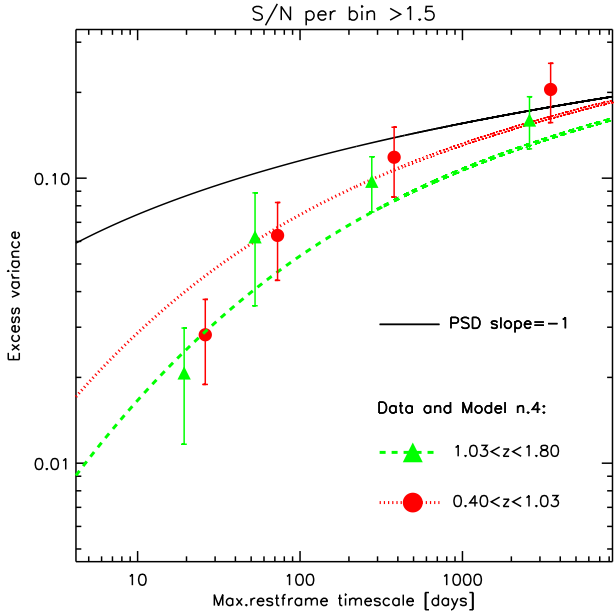}
     \caption{Excess variance vs maximum sampled timescale for sources in the bright-R sample with S/N per bin $>1.5$, in 2 redshift ranges. The solid black line indicates the model excess variance in the case when PSD$(\nu)\propto \nu^{-1}$. The discontinuous lines represent instead the prediction from our best-fit bending power-law model (Model 4, \S \ref{modelingagn}). }
       \label{var_vs_timescale}
\end{figure}

Since the sampled timescales correspond to different rest-frame timescales at different redshifts, we grouped our sources (up to $z\sim 2$) in the 2 redshift intervals listed in the legend of Figure \ref{var_vs_timescale}. They were defined in such a way that the interval width was kept as small as possible to reduce the internal difference in rest-frame timescales, and at the same time there were at least 15 sources in each bin. The average $\sigma^2_{NXS}$ of all the sources in each redshift bin is shown in Figure~\ref{var_vs_timescale}, as a function of the maximum rest-frame timescale (estimated at the mean $z$ of each bin).

At each timescale, the higher redshift measurements are systematically smaller than the average variability amplitude in the lower redshift bins. This is due to the different luminosity ranges sampled at each redshift. However, the important result is that, for both redshift bins, the variability amplitude clearly decreases toward shorter rest-frame timescales. Although the data plotted in Figure~\ref{var_vs_timescale} are not directly PSD measurements (since the excess variance estimates the integral of the PSD between the minimum and maximum sampled timescale),
the decrease of the excess variance with decreasing time scale is 
direct observational evidence for the red-noise nature of the variability process of the high-redshift AGNs. 


To demonstrate that this is indeed the case, in Figure~\ref{var_vs_timescale} we overplot a model prediction based on the assumption that the average intrinsic PSD has a power-law shape. The black solid line shows $\sigma_{mod}^2$ when the PSD is a single power-law  of the form PSD$(\nu) = A \nu^{-1}$ (this model is appropriate for local  AGNs on long timescales, e.g. \citealt{Uttley02, Markowitz03, McHardy04, McHardy06}). In this case $\sigma_{mod}^2 = A\ln\left(\frac{\nu_{max}}{\nu_{min}}\right )$ where $\nu_{min}$ and $\nu_{max}$ are the lowest and highest sampled rest-frame frequencies. We fixed $\nu_{max}=(1+z)/(86400~\Delta t_{min}^{obs})$ s$^{-1}$ where $\Delta t_{min}^{obs}=0.25$ d\footnote{Using a value of 0.95 d, appropriate for the intermediate timescales, has negligible effects as most of the variability power is concentrated anyway on the longest timescales.}, using the average redshift of the low-$z$ sample. We also chose $A$ so that the model excess variance matches the excess variance of the longest timescale for the low-redshift bin, in order to display the model behaviour.

Qualitatively, the model predictions (decrease of excess variance with decreasing $t$) are similar to what we observe. However, the model has a shallower slope than the observed one. This suggests that such a flat PSD, typical of local AGNs on long timescales, is inadequate to describe the measured excess variance on short timescales. A bending power-law model with a high-frequency cutoff, described in detail in \S \ref{modelingagn}, does a much better job in reproducing the observed trend. We believe that Figure~\ref{var_vs_timescale} not only demonstrates that the high-redshift AGNs have PSDs which are well represented, on average, by a power-law, but also that their PSD ``breaks" above some characteristic frequency, as observed in several nearby AGNs.

\section{Testing variability models and tracing the AGN accretion history}
\label{accretion_section}

The discussion above demonstrates that it is difficult to draw firm conclusions about the  dependence of the variability amplitude on the underlying AGN physical parameters and its evolution, based on the excess-variance vs luminosity/redshift/timescale plots. However,  with the CDF-S data, we can now study more accurately the $\sigma_{NXS}^2-L_X$ relation at different redshifts, by treating properly the differences in sampled luminosities and timescales (due to differences in $z$). To this end we divided the bright-R sample sources in four redshift intervals and we computed the average $\sigma^2_{NXS}$ of sources in luminosity bins containing at least 15 sources. We considered the two redshift bins that we also considered in \S\ref{var_timescale}, and to increase the redshift range we also considered the [1.8 - 2.75] and [2.75 - 4] redshift bins \footnote{ Due to the small number of sources at $z>2.75$ we grouped all sources in one bin containing 10 objects.}.

\begin{figure*}
   \centering

   \includegraphics[width=0.24\textwidth]{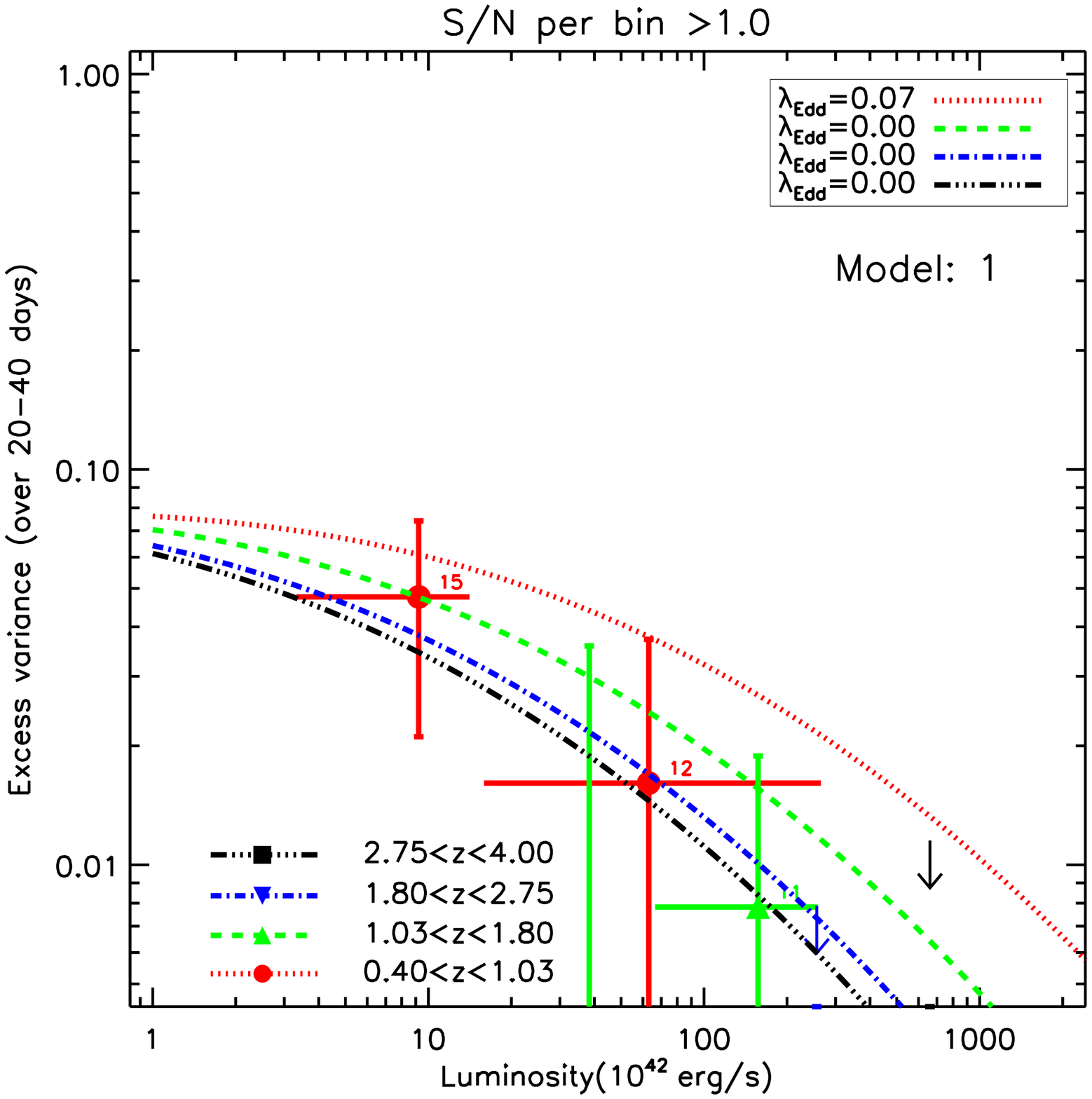}
    \includegraphics[width=0.24\textwidth]{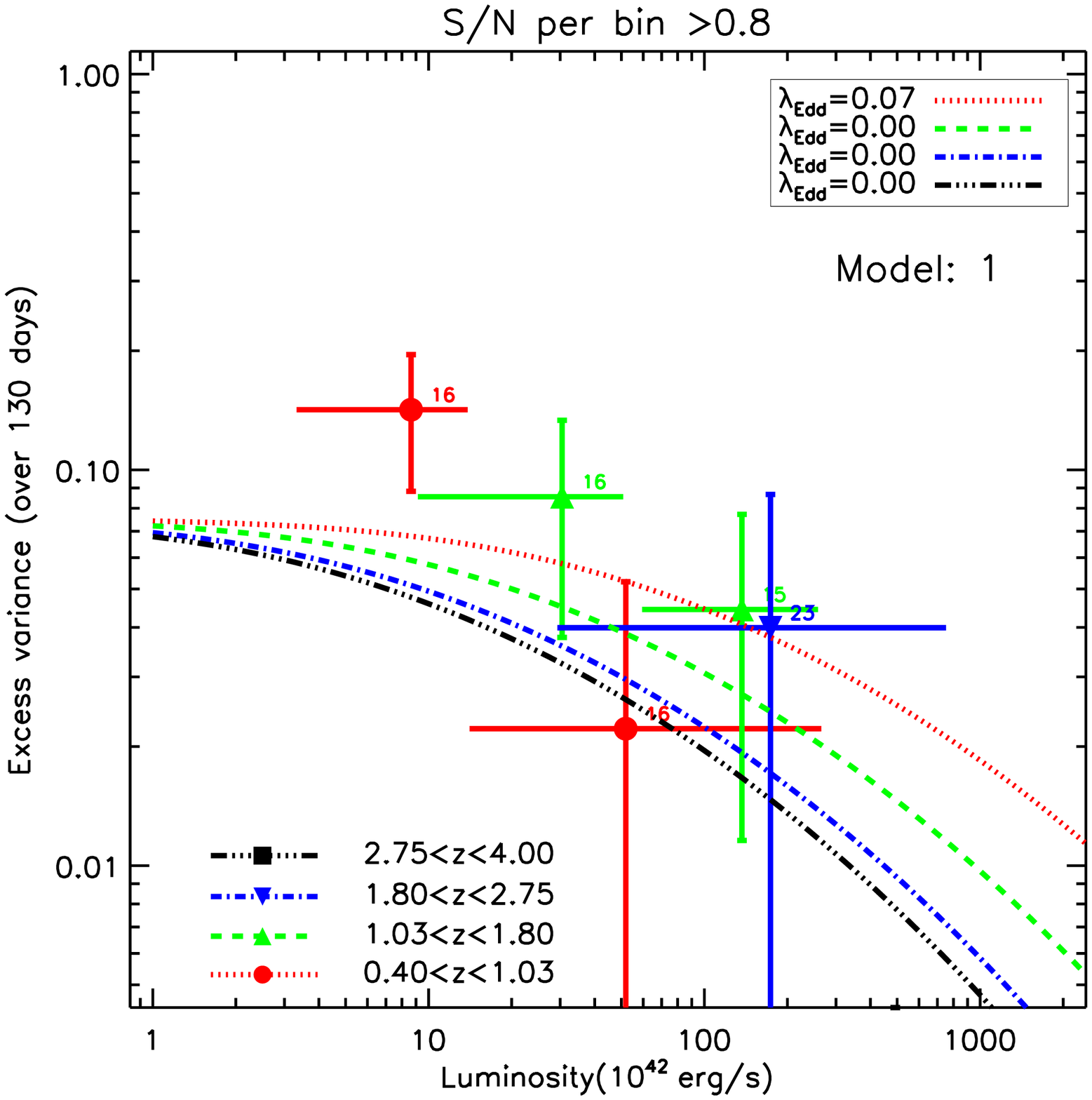}
    \includegraphics[width=0.24\textwidth]{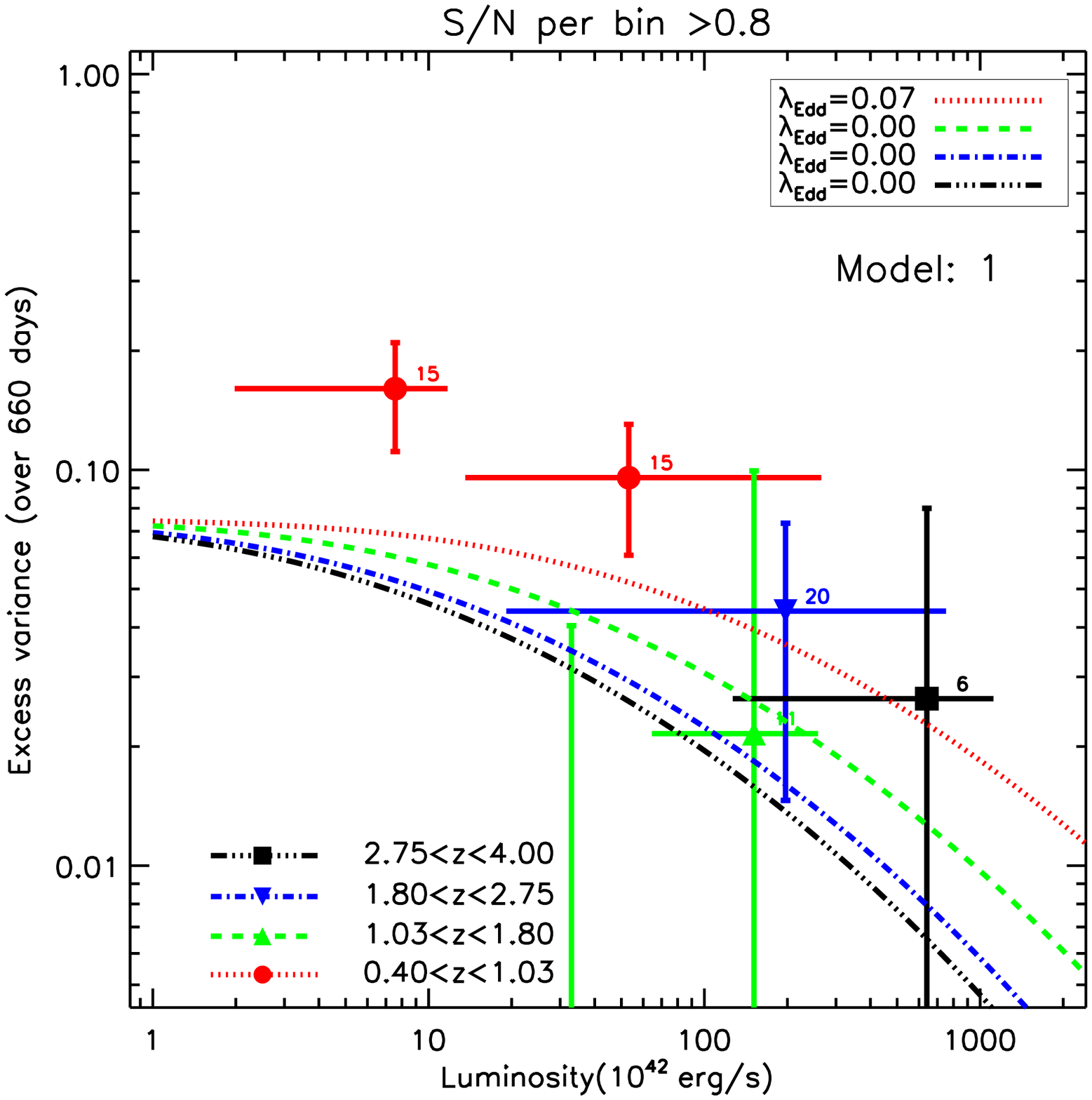}
    \includegraphics[width=0.24\textwidth]{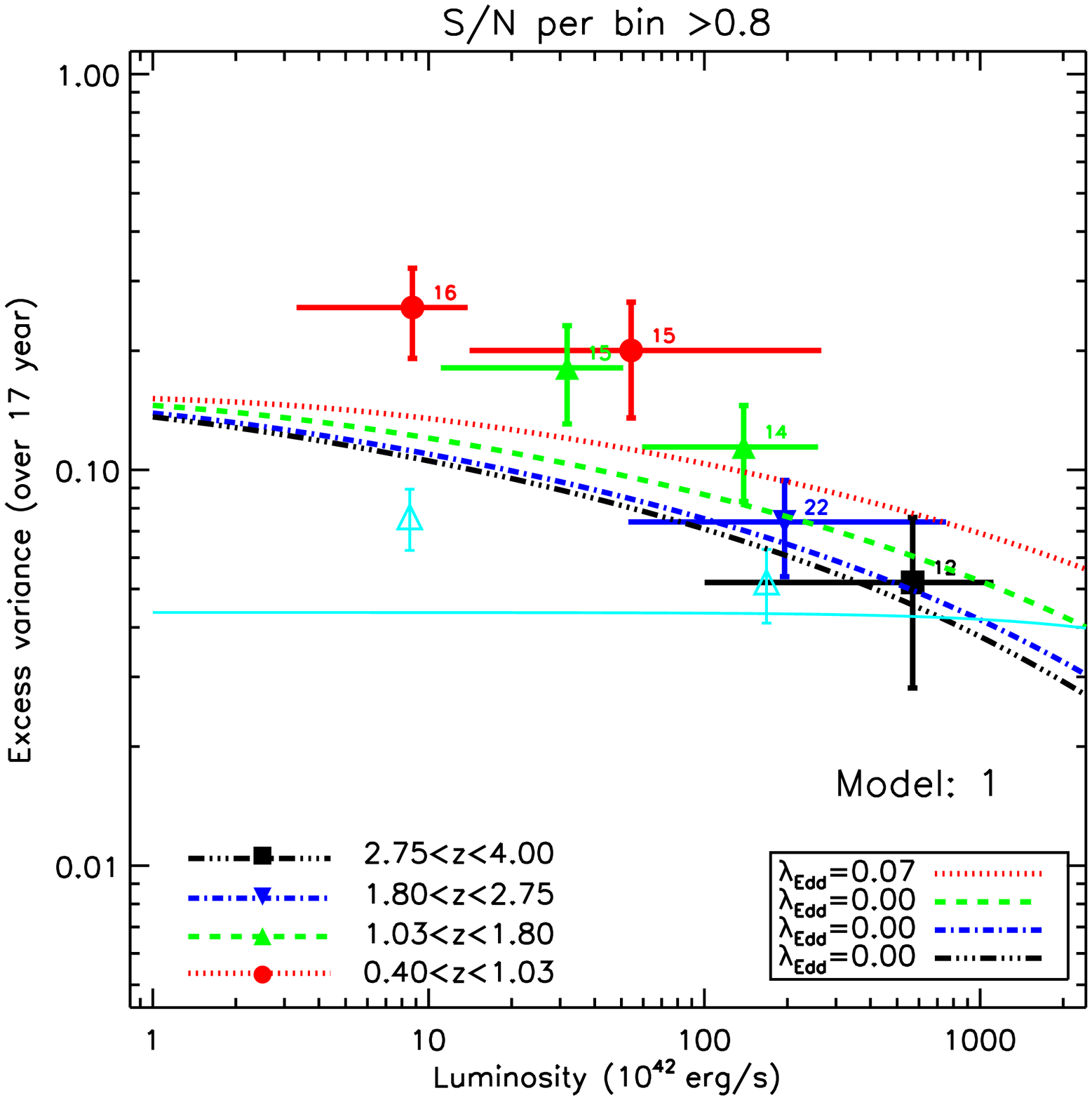}

   \includegraphics[width=0.24\textwidth]{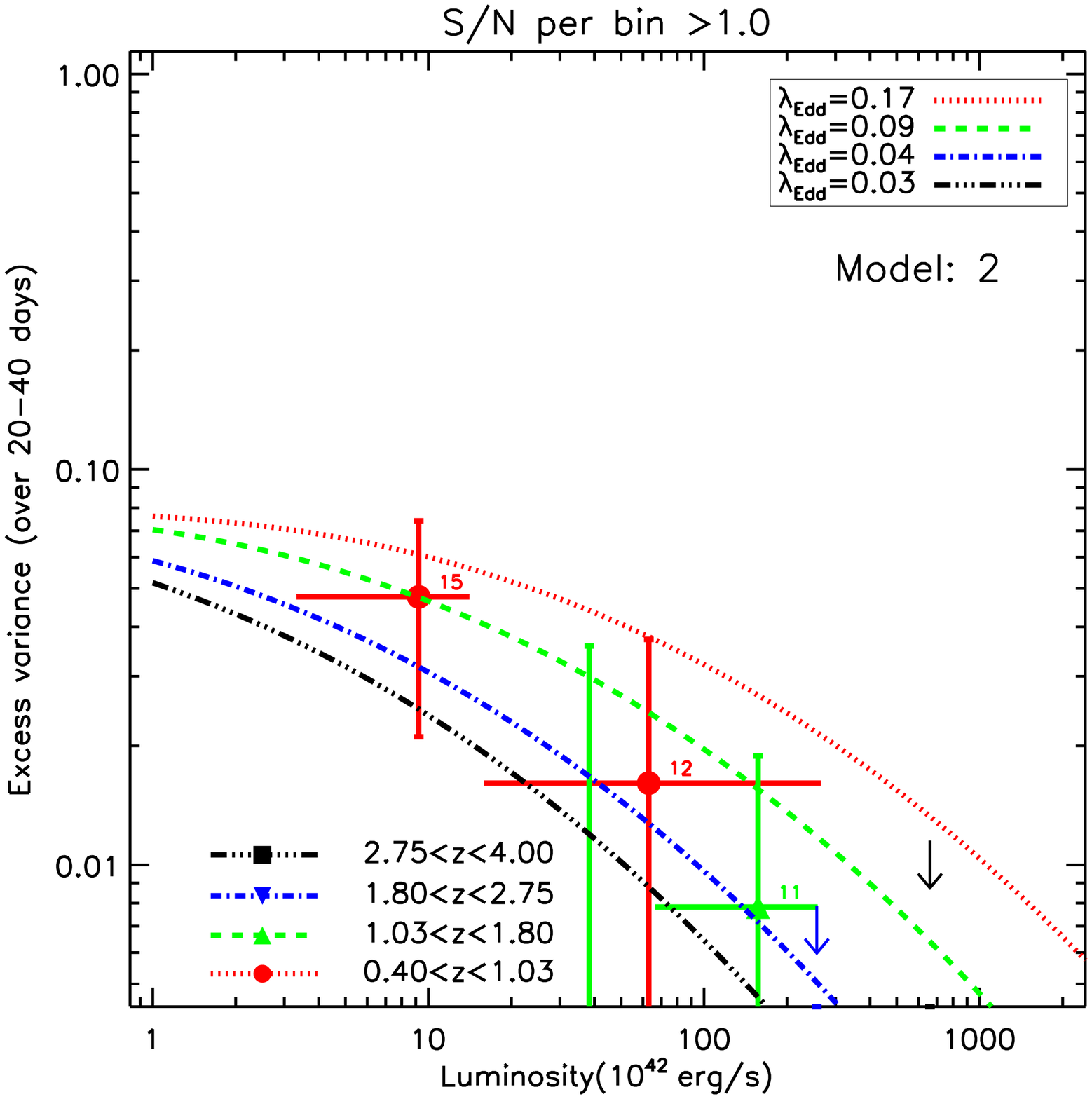}
    \includegraphics[width=0.24\textwidth]{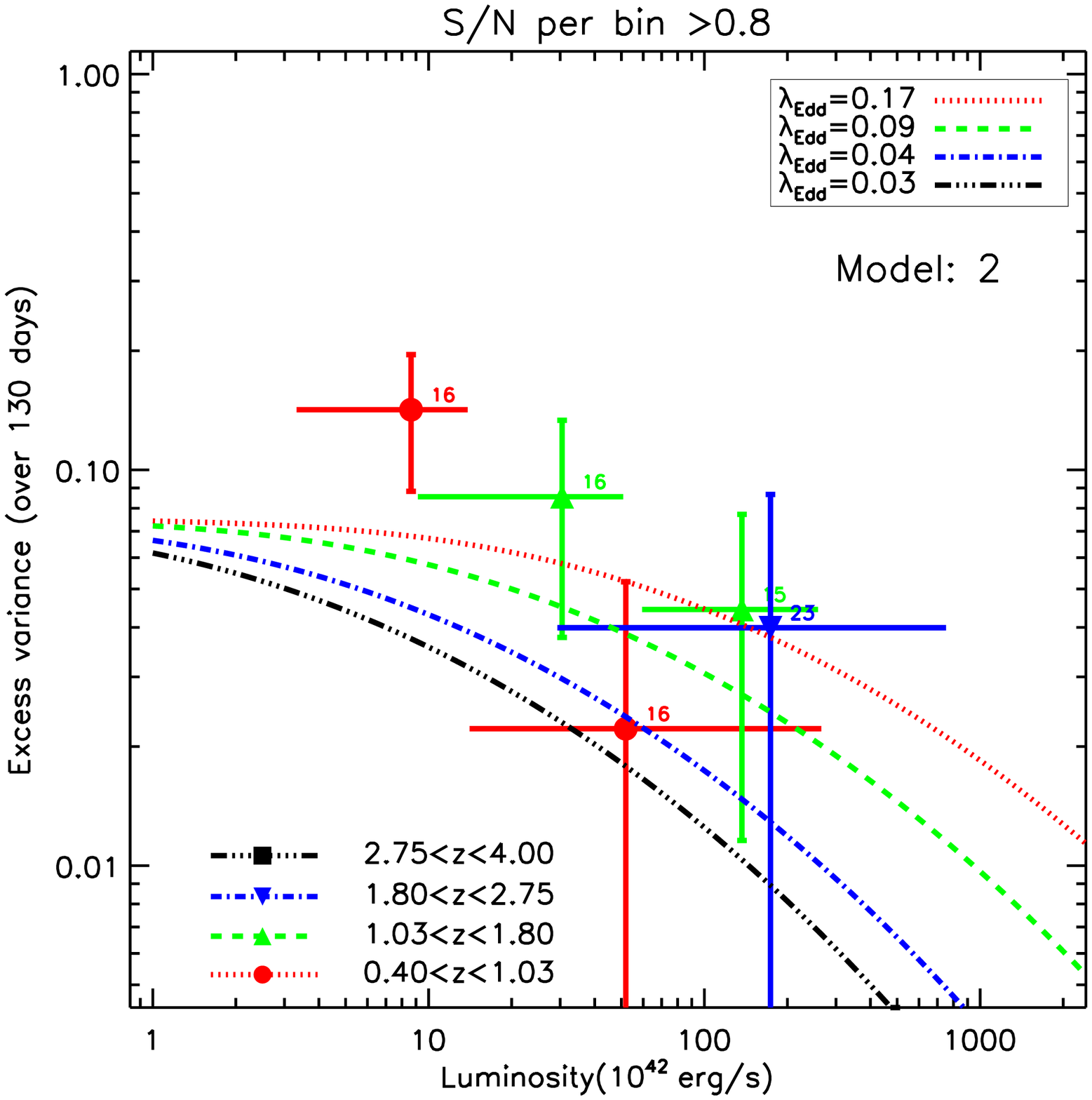}
    \includegraphics[width=0.24\textwidth]{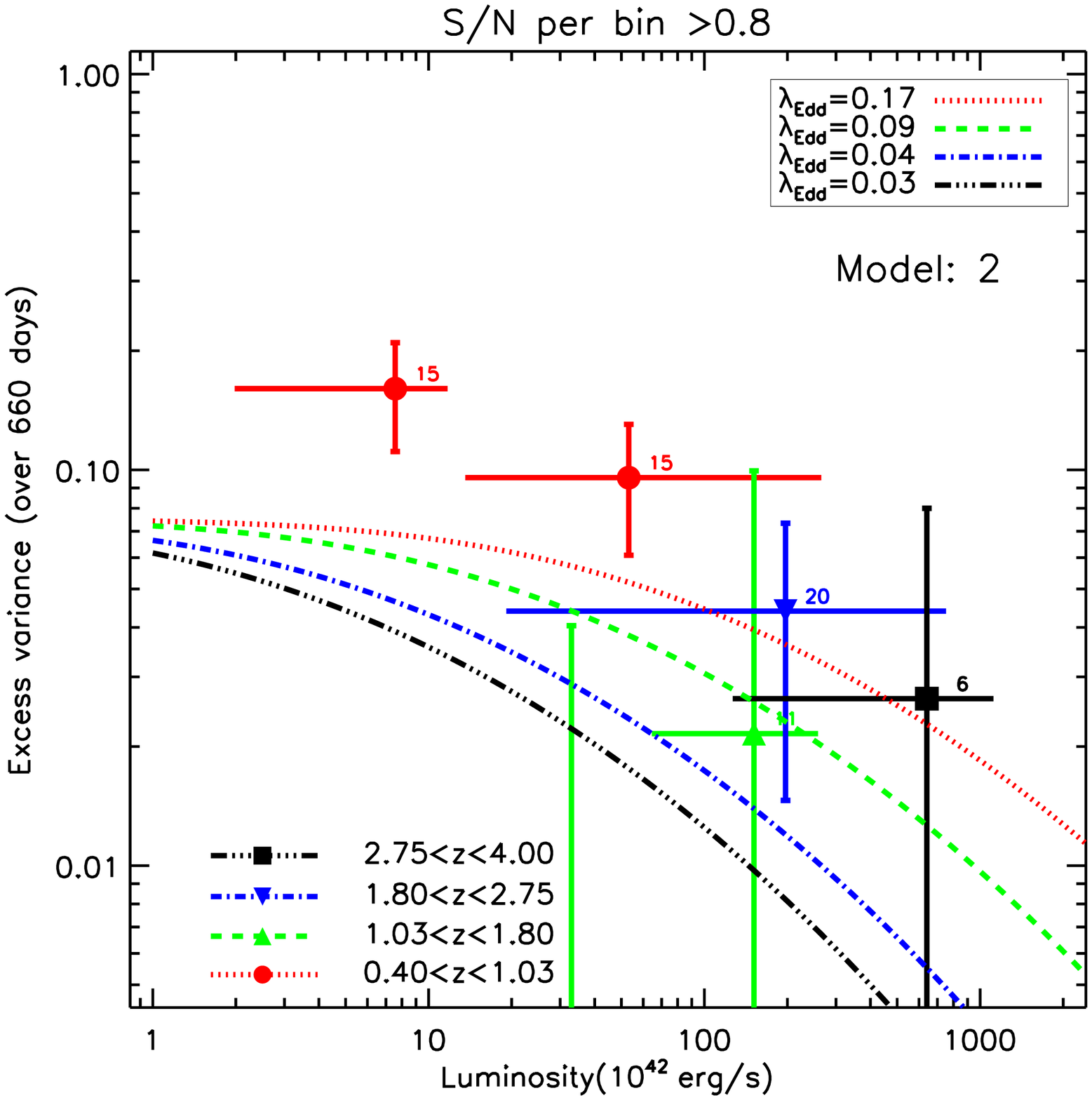}
    \includegraphics[width=0.24\textwidth]{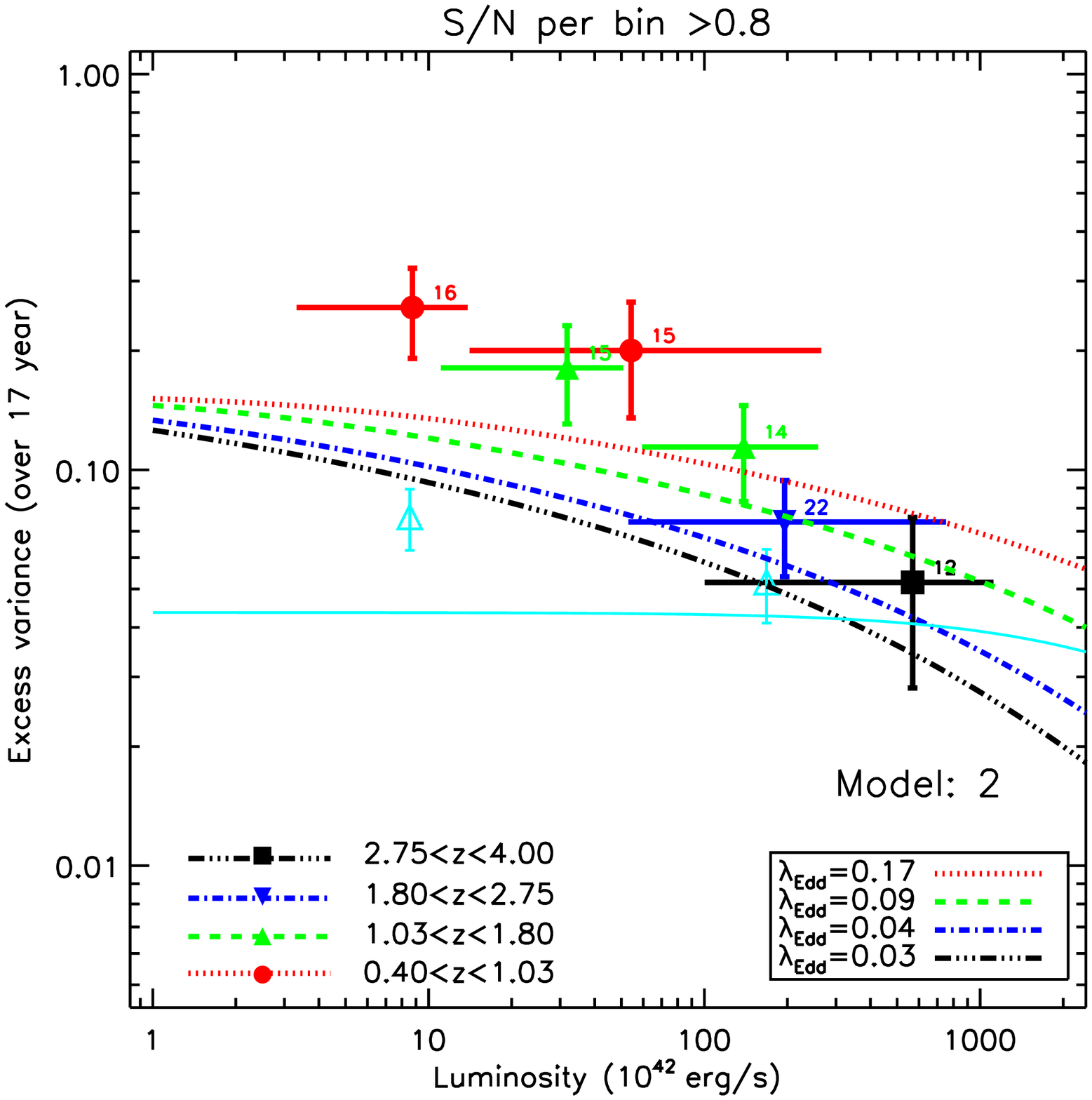}

   \includegraphics[width=0.24\textwidth]{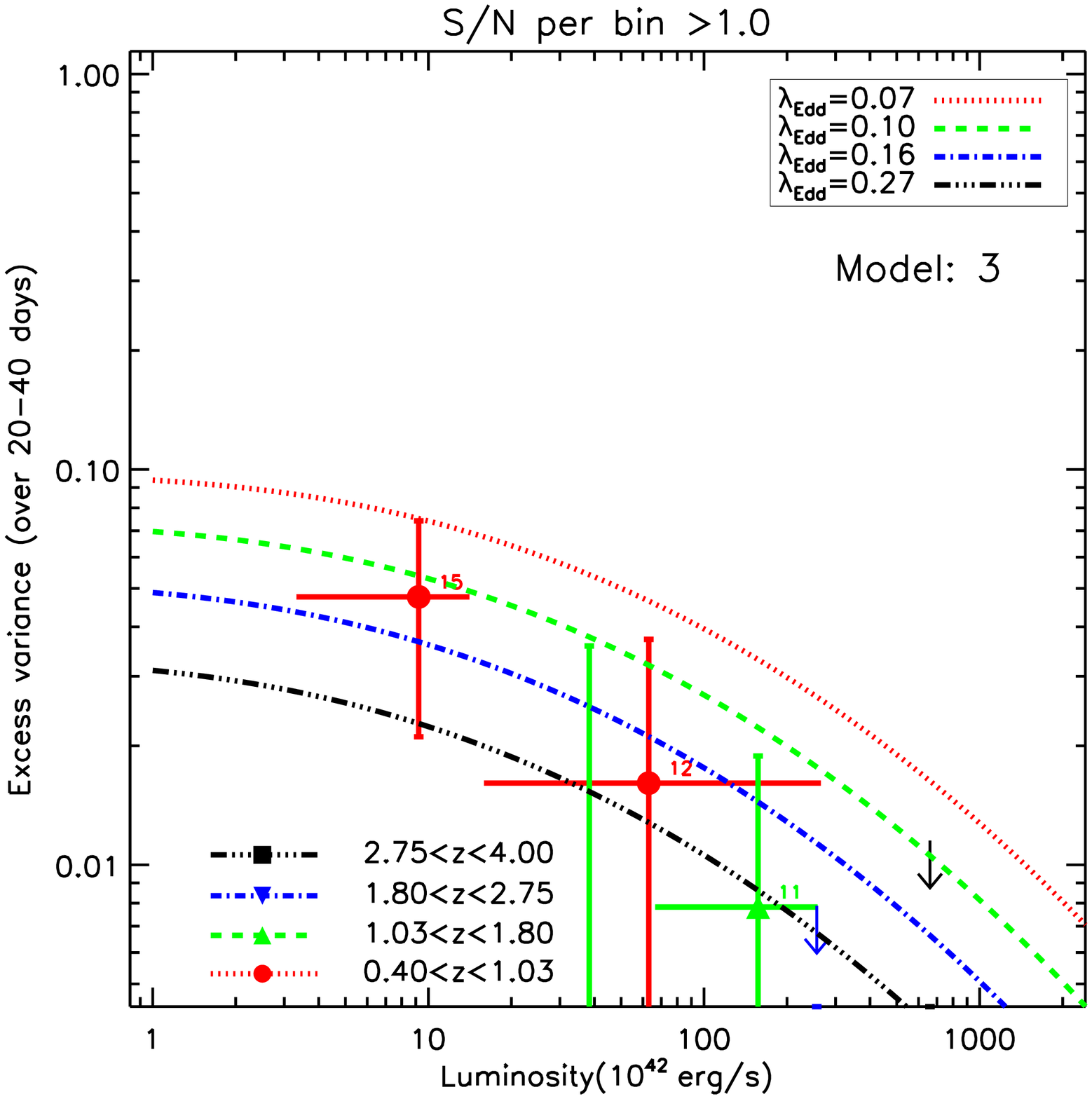}
    \includegraphics[width=0.24\textwidth]{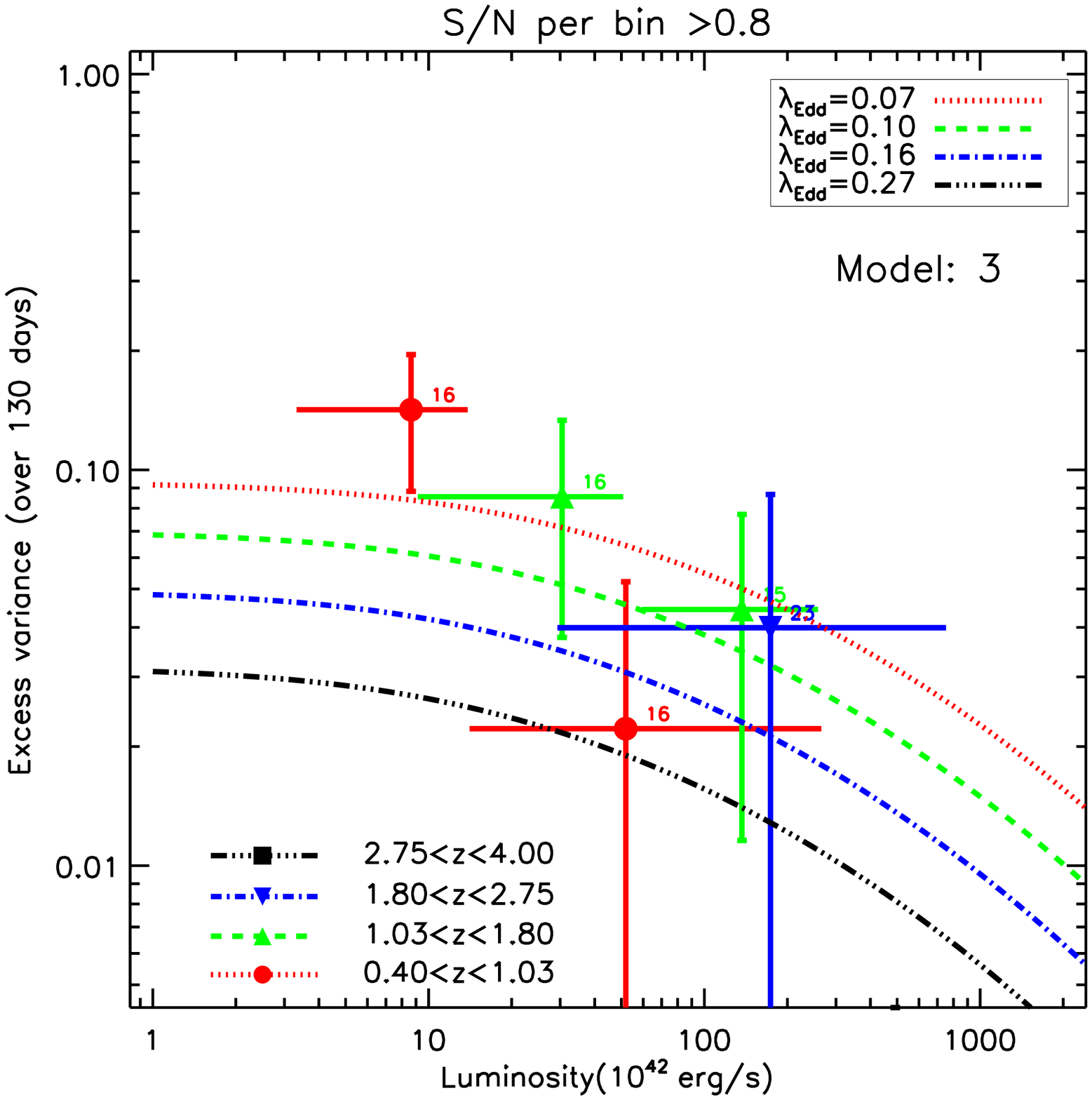}
    \includegraphics[width=0.24\textwidth]{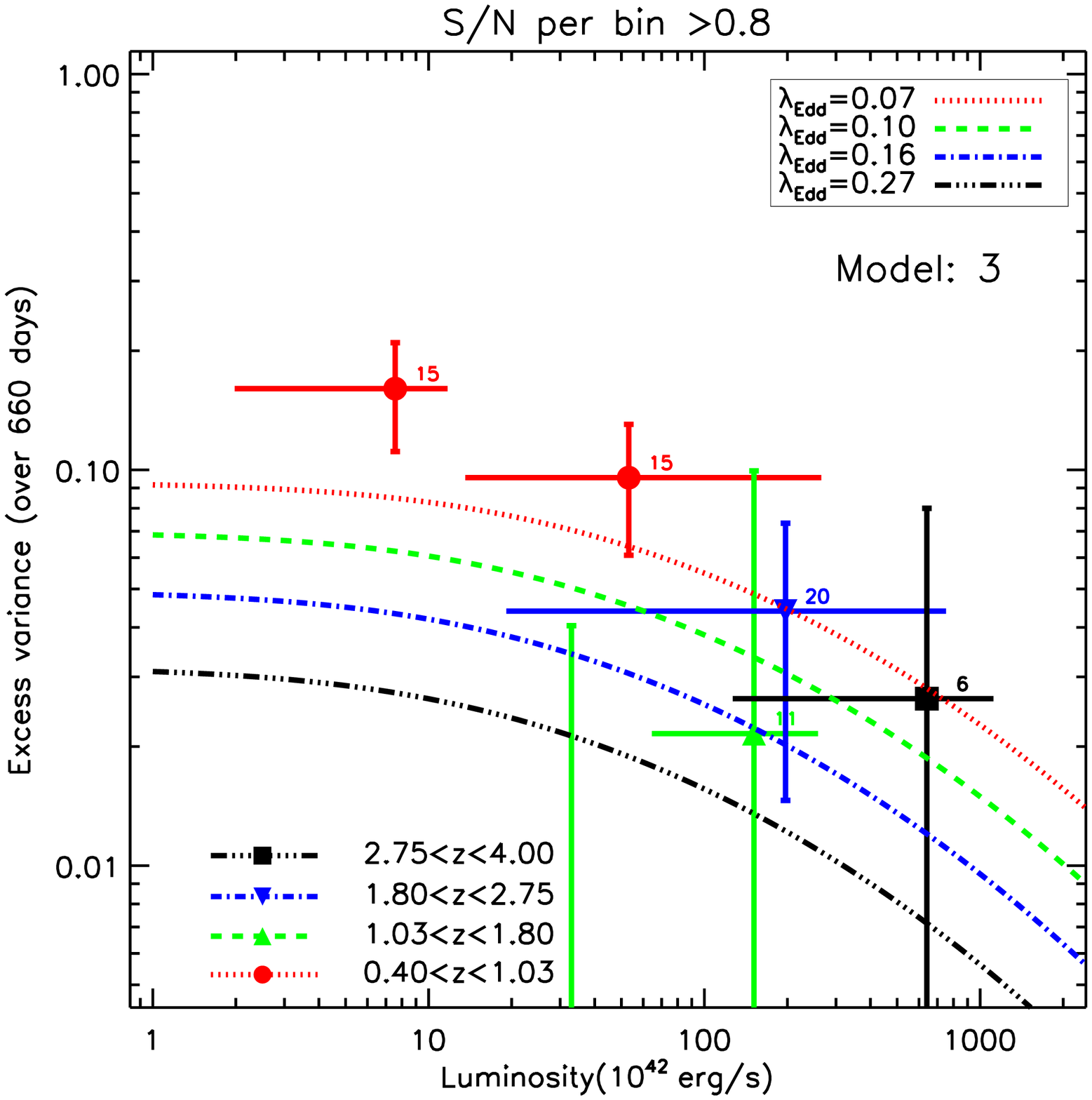}
    \includegraphics[width=0.24\textwidth]{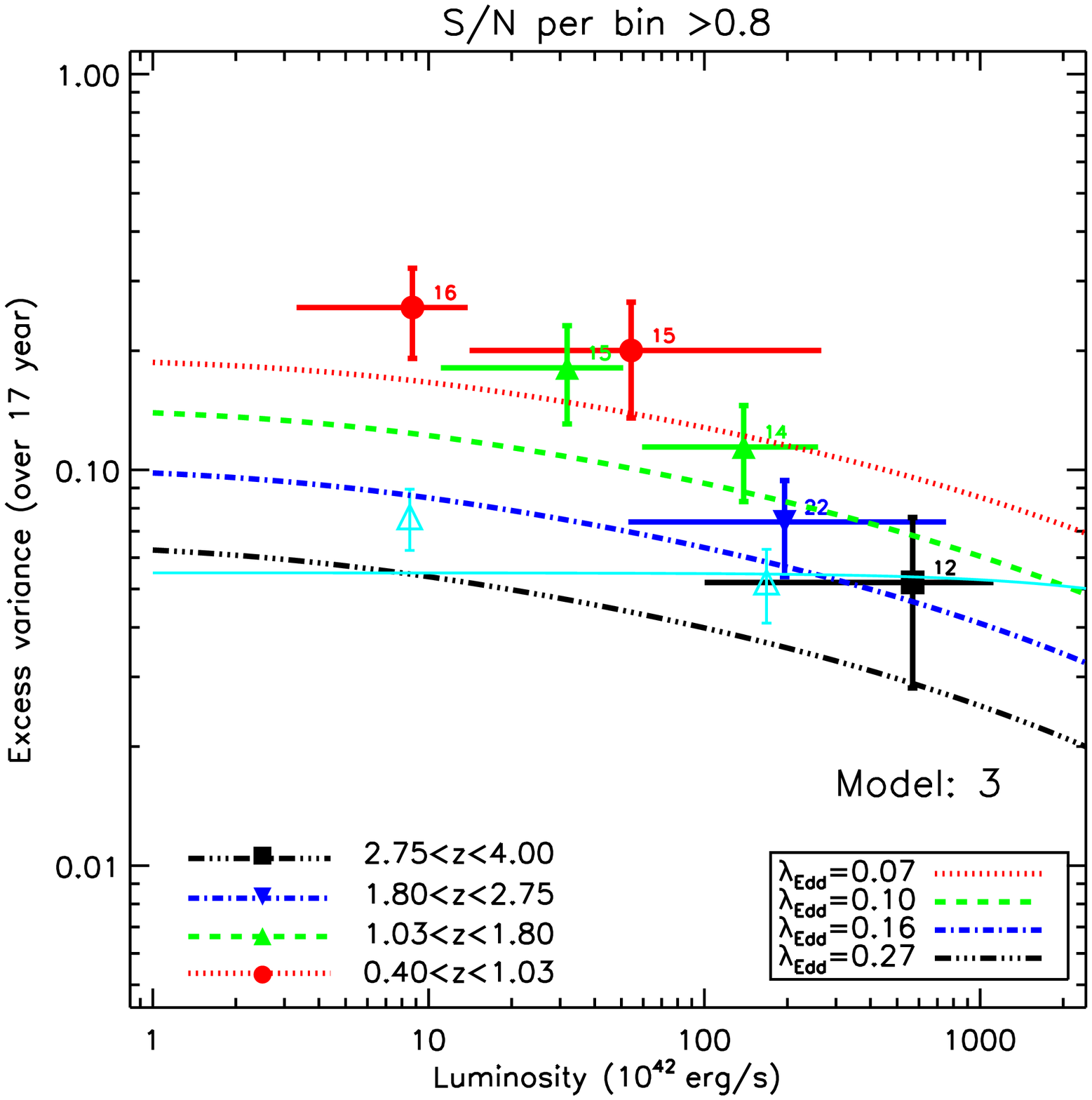}

   \includegraphics[width=0.24\textwidth]{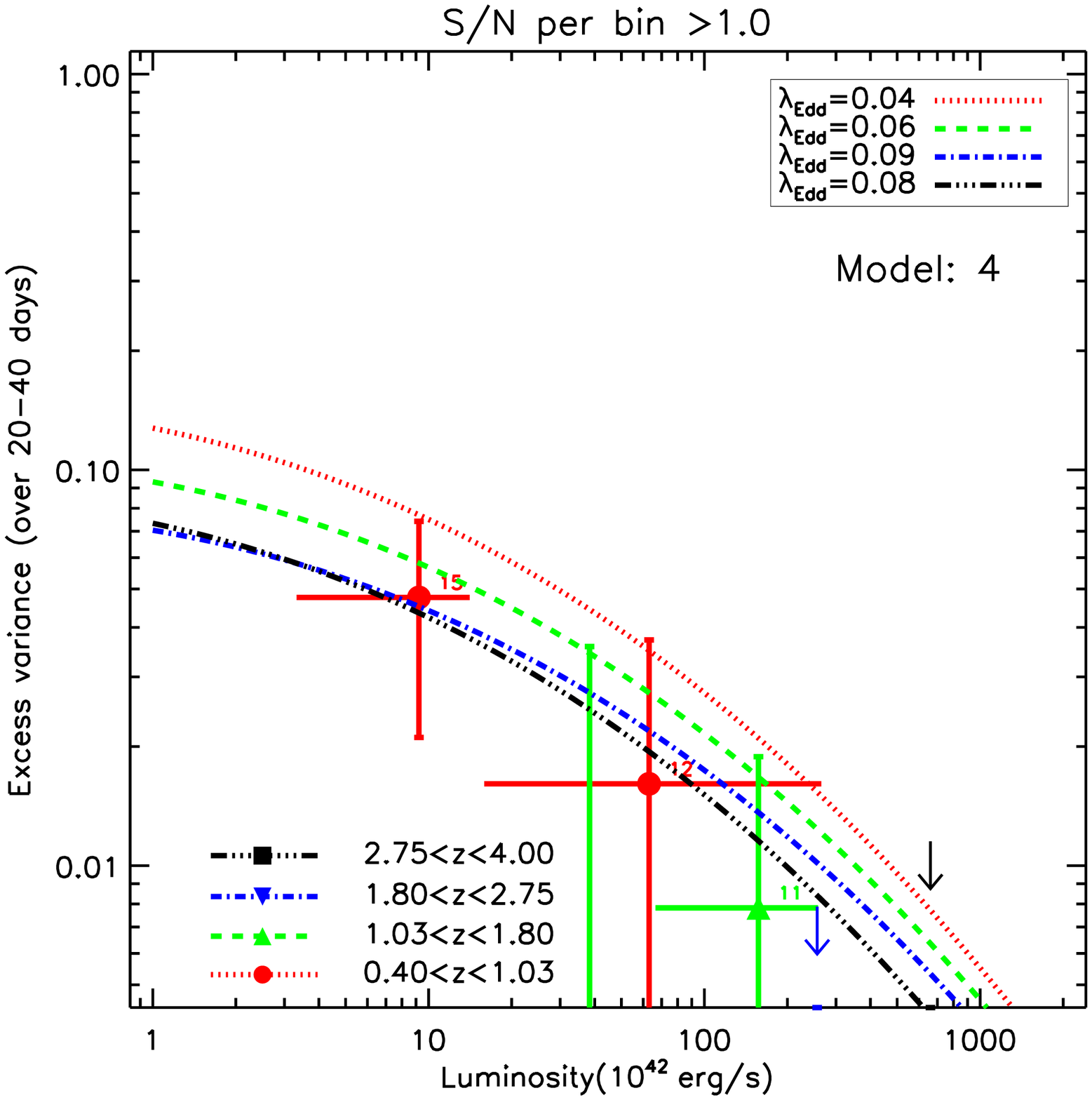}
    \includegraphics[width=0.24\textwidth]{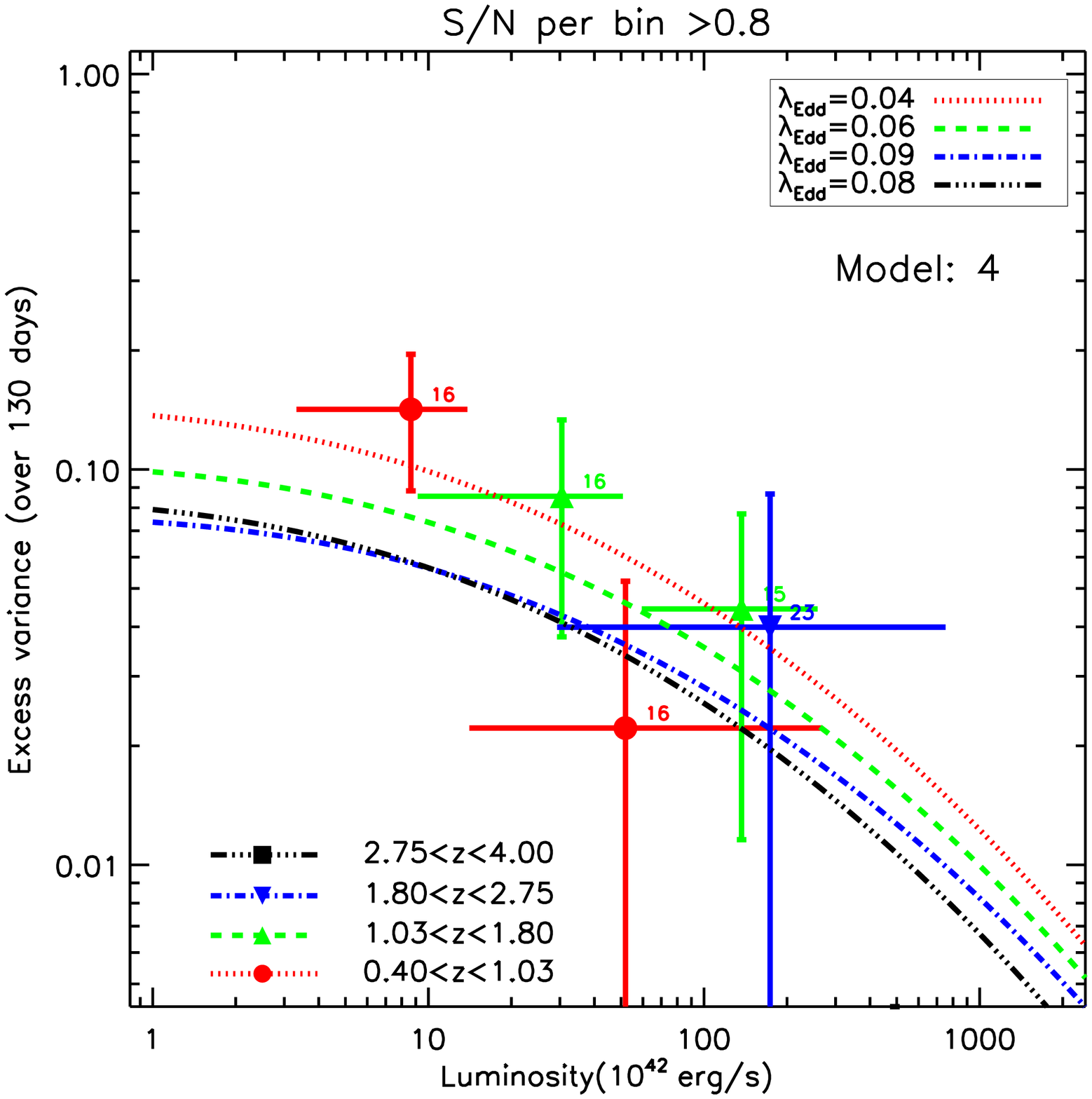}
    \includegraphics[width=0.24\textwidth]{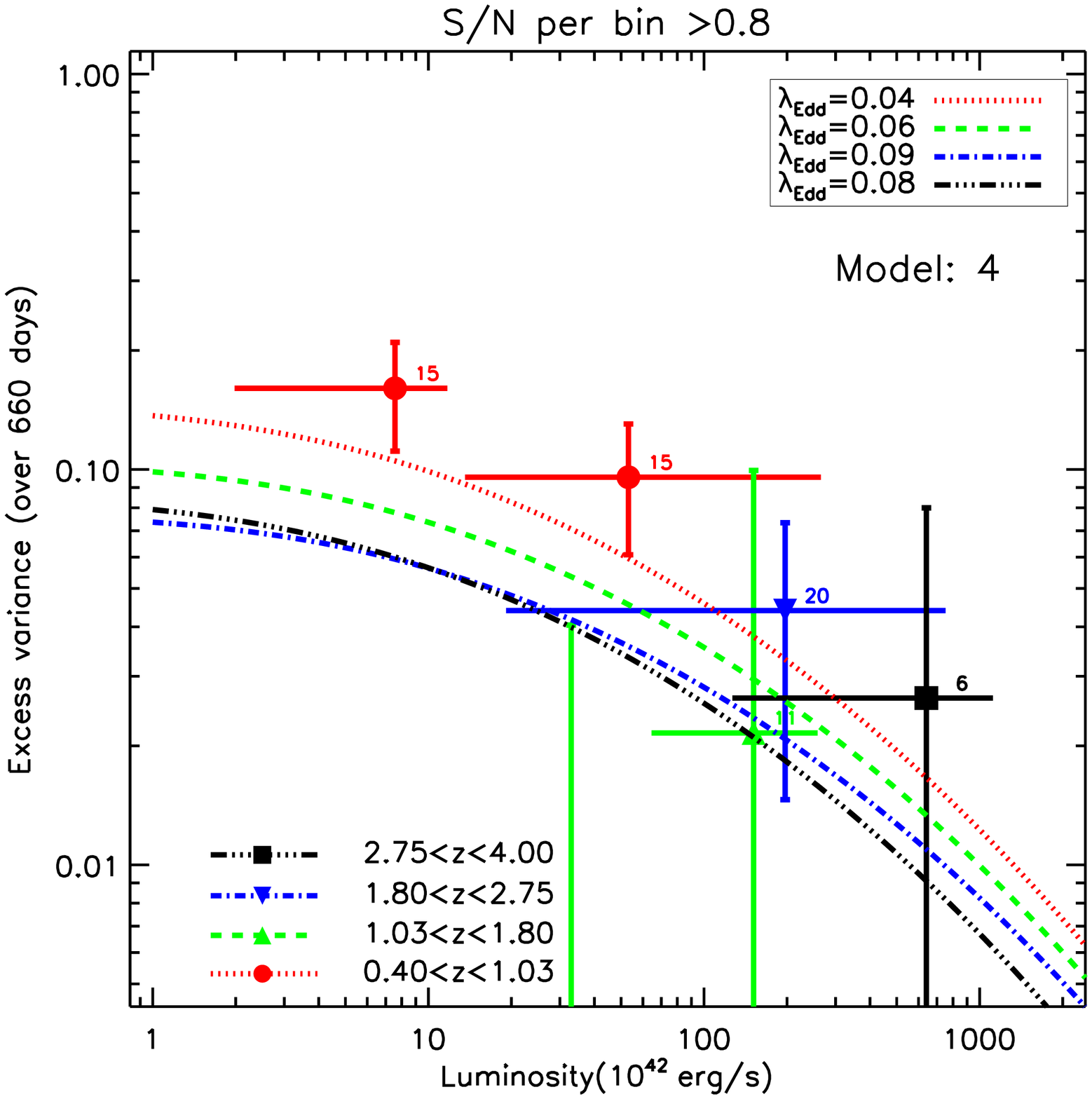}
    \includegraphics[width=0.24\textwidth]{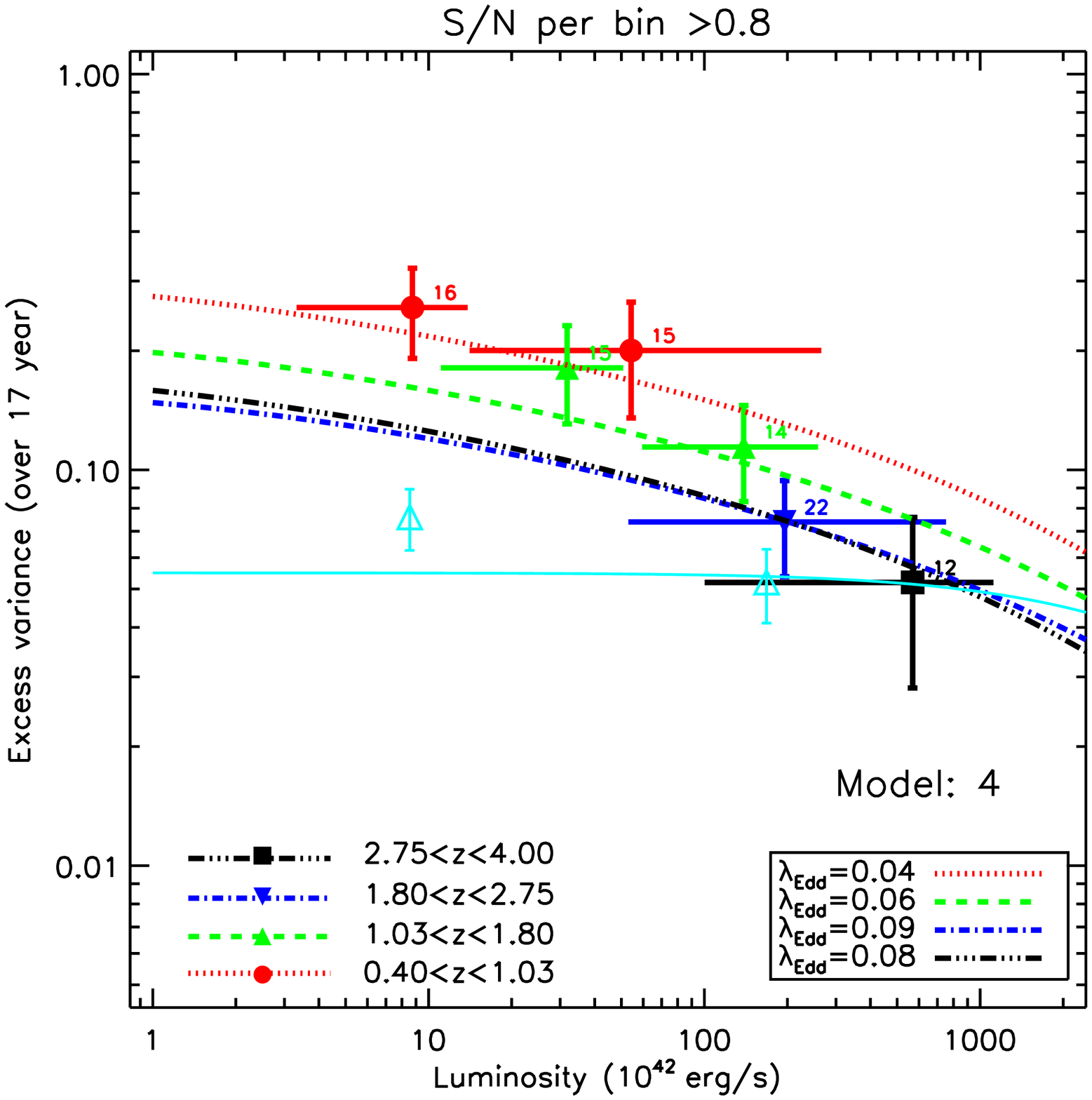}
\caption{The $\sigma^2_{NXS}$ vs $L_X$ relation for the Bright-R sources in different redshift intervals. The four columns correspond to the different timescales discussed in \S \ref{var_timescale}. The CDF-S data are binned in groups of $> 15$ sources (the exact number is printed above each point). The arrow's in the panels of the first column represent the $3\sigma$ upper limit. The lines in each panel show the predictions of Models 1, 2, 3 and 4 with parameters set to the best-fit solutions reported in Table \ref{fit_varacc_table} (which allow a variable Eddington ratio as a function of redshift).  The open triangles and solid cyan line in the rightmost panels represent the local AGN data from \citet{Zhang11} (each point includes 14 sources) and the corresponding model prediction.}
         \label{model_comparison_varacc}
\end{figure*}

The four columns of Figure~\ref{model_comparison_varacc} show the $\sigma^2_{NXS}$ measurements in the four different redshift intervals plotted as a function of X-ray luminosity, for the four different timescales  discussed in \S \ref{var_timescale}.  
The decrease in variability with increasing X-ray luminosity is confirmed on most timescales and redshift bins, at least up to $z\sim 2$ where we probe a large enough range of luminosities.
Arguably, the uncertainty on the individual points is large, but we do not observe any significant increase of variability amplitude with redshift, in any of the timescales we considered. The amplitude of the $\sigma^2_{NXS}-L_{X}$ relations increases with increasing timescale, which is caused by the red-noise character of the observed variations. On the shortest timescales, the anti-correlation between $\sigma^2_{NXS}$ and $L_{X}$  is steep. Then it flattens as we sample increasingly longer time intervals; this behaviour is in agreement with the scenario where the intrinsic PSD is represented by a bending power-law with a high-frequency cutoff.

In order to to understand  the observed complex dependence of the variability-luminosity relation in various redshift bins and over different timescales, we fitted the data shown in Figure~\ref{model_comparison_varacc}  with predictions of PSD models, which are frequently used to parametrize the observed power-spectra of nearby, X--ray bright AGN, as we explain below. 
To constrain better the models at the lowest redshifts, which are not sampled by the CDF-S population, we also considered the data from the sample of local AGNs studied by \citet{Zhang11}. The \citet{Zhang11} lightcurves are based on 14-year long \textit{RXTE} monitoring campaigns, closely matching the full 7 Ms observed timescales. Note that the \textit{RXTE} monitoring cadence only allows us to probe the longest timescales, with no equivalent to the additional long, intermediate and short timescales probed for the CDF-S sources. Therefore, the \citet{Zhang11} data are thus only shown in the rightmost panels of Figure~\ref{model_comparison_varacc}.

\subsection{Modeling the AGN variability}
\label{modelingagn}

The AGN PSDs have been modelled in the past by either a simple power-law, or a broken or bending power-law \citep[see e.g.][]{Markowitz03}, where the normalization and the position of the break depend on the AGN physical parameters such as BH mass and accretion rate. Here we adopt the bending power-law model.  Following \cite{McHardy04b, Gonzalez-Martin2012}, the PSD is represented by the function:

\begin{equation}
\label{PSDmod}
\mbox{PSD}(\nu)=A\nu^{-1}\left ( 1+\frac{\nu}{\nu_b} \right ) ^{-1}
\end{equation}

\noindent where $A$ is the normalization factor and $\nu_b$ is the break (or bending) frequency; the PSD thus has a logarithmic slope of -1 for $\nu<<\nu_b$ which becomes -2 for $\nu>>\nu_b$. The model PSD we adopt here is based on PSD studies of local AGNs. In principle, variability analysis of high-redshift AGNs should also test if these models are appropriate for the modelling of their X--ray variability properties. The estimation of the PSD of high-redshift AGNs is challenging, mainly due to the poor temporal sampling of the existing light curves. However, as we have argued in 
\S\ref{var_timescale}, the results plotted in Figure~\ref{var_vs_timescale} already suggest that PSD models like the ones defined above are appropriate to describe the X--ray variability of the high redshift AGN. 

According to the model PSD, the lightcurve variance  takes the form:

\begin{equation}
\label{ltcvar}
\begin{aligned}
\sigma_{mod}^2 ={} & \int^{\nu_{max}}_{\nu_{min}}\mbox{PSD}(\nu)\ d\nu=\\
            ={} & A\left [ \ln\left(\frac{\nu_{max}}{\nu_{min}}\right )-\ln\left(\frac{\nu_b+\nu_{max}}{\nu_b+\nu_{min}}\right ) \right ], \\
\end{aligned}
\end{equation}

\noindent where $\nu_{max}$ and $\nu_{min}$ are the highest and lowest rest-frame frequencies sampled by our lightcurves. In particular $$\nu_{min}=(1+z)/\Delta t_{max}^{obs}; ~~\nu_{max}=(1+z)/\Delta t_{min}^{obs}$$ where $\Delta t_{max}^{obs}$ is the total duration of the lightcurve and $\Delta t_{min}^{obs}$ is the minimum sampled timescale. In our case $\Delta t_{max}^{obs}=(45,127,654,6005)~\mbox{days}$ for the short, intermediate, long and 7 Ms timescales, respectively. In the case of unevely sampled lightcurves, the choice of $\Delta t_{min}^{obs}$ is not obvious. We choose the minimum gap between consecutive observation in our lightcurve, corresponding to $\Delta t_{min}^{obs}=(0.25,0.95,0.25,0.25)~\mbox{days},$ for the short, intermediate, long and 7 Ms timescales respectively.

To link the variability of the AGN to its physical properties we explore four different variations of the PSD model defined by eq.\,~\ref{PSDmod}: 

\begin{itemize}
\setlength\itemsep{1em}
\item{\textbf{Model 1:}} the PSD amplitude at $\nu_b$ is constant, in terms of $\nu_b\times \mbox{PSD}(\nu_b)= 0.02$, for all AGNs as suggested by \citet[see also \citealt{Gonzalez-Martin2011}]{Papadakis04}, while the break frequency depends on the BH mass as $\nu_b=580/(M_{BH}/M_\odot)$ s$^{-1}$ as suggested by \cite{Gonzalez-Martin2012}; 
\item{\textbf{Model 2:}} the PSD amplitude is constant as in Model 1, while the break frequency depends both on the BH mass and the accretion rate so that 
$$\nu_b=(200/86400) (L_{44, bol}) (M_{6, BH})^{-2}~\mbox{s}^{-1}$$  
as in the prescription of \cite{McHardy06}, where $L_{44, bol}$ is the bolometric luminosity in units of $10^{44} \mbox{erg s}^{-1}$ calculated applying the recipe of \citet{Lusso12}\footnote{We verified that using different recipes, such as the one proposed by \cite{Marconi04} we obtain consistent results.}, and $M_{6, BH}$ is the BH mass in units of $10^6 M_\odot$;
\item{\textbf{Model 3:}} similar to Model 1, where $\nu_b=580/(M_{BH}/M_\odot)$ s$^{-1}$, but the PSD normalization is itself dependent on the accretion rate as $$\nu_b\times \mbox{PSD}(\nu_b)=3\times 10^{-3}\lambda_{Edd}^{-0.8}$$ (with $\lambda_{Edd}=\frac{\dot{m}}{\dot{m}_{Edd}}$) as proposed by \cite{Ponti12};  
\item{\textbf{Model 4:}} similar to Model 2, where $\nu_b=(200/86400) (L_{44, bol}) (M_{6, BH})^{-2}~\mbox{s}^{-1}$, but the PSD normalization depends on the accretion rate as in Model 3.
 \end{itemize}

\noindent In summary,  the first two models assume that only the break frequency depends on the AGN physical parameters, i.e. the BH mass (Model 1) or BH mass and accretion rate (Model 2). The last two models include a dependence of the PSD amplitude on the accretion rate as well.  

\subsection{Model fit procedure}
\label{fit_proced}

We fit each one of the four models presented in \S \ref{modelingagn} to the data points plotted in Figure~\ref{model_comparison_varacc}, over all luminosities, timescales and redshifts simultaneously. Equation (\ref{ltcvar}) was integrated over the range of rest-frame frequencies sampled in each redshift interval, as described in \S \ref{modelingagn}, in order to derive the excess variance $\sigma^2_{mod}$.
The observed $\sigma^2_{NXS}$ is an estimator of the intrinsic lightcurve variance $\sigma_{mod}^2$ provided that we take into account the biases introduced by the sampling pattern (red-noise leakage and uneven sampling). We adopted the \citet{Allevato13} recipe to correct for such biases, deriving the predicted excess variance as $\sigma_{pred}^2= \sigma^2_{mod} / (C \cdot 0.48^{\beta-1})$, where $\beta$ is the PSD slope below the minimum sampled frequency $\nu_{min}$, and $C$ is a corrective factor dependent on the sampling pattern. In our models $\beta$ was estimated as the average slope from $\nu=\nu_{min}/5$ and $\nu_{min}$, i.e. the range from where most of the leaking power originates \citep[cf.][]{Allevato13}. The factor $C$ ranges from 1 to 1.3  and allows us to account for the missing power due to the gaps in the lightcurve; we adopted the upper value of $C=1.3$ adequate for the sparse sampling of CDF-S lightcurves and when $\beta=1.0-1.5$ but we verified that changing the bias factor yields consistent results within $\Delta \lambda_{Edd}\sim 0.03$, where higher accretion rates correspond to lower bias values.

The only free parameter in our models is the average accretion rate $\lambda_{Edd}$. For a given accretion rate we compute the bolometric luminosity $L_{bol}$, using X-ray luminosity value of each point, according to the \citet{Lusso12} prescription. Using $\lambda_{Edd}$ and $L_{bol}$, we then compute the average BH mass for all AGNs which contributed to $\sigma_{NXS}^2$ in each bin, and $\nu_b$ and the normalization $A$ according to each model prescription. Knowing $\nu_b$ and  $A$ is then possible to compute $\sigma_{mod}^2$ using equation (\ref{ltcvar}) for each point in the $\sigma_{NXS}^2-L_X$ relations.
The best-fit $\lambda_{Edd}$ is then found by $\chi^2$ minimization of the differences between $\sigma_{NXS}^2$ and $\sigma_{pred}^2$ (we note again that we used $\sigma_{pred}^2$ and not $\sigma_{mod}^2$, to take into account the bias due to sampling and red noise leak).

We note that the CDF-S data used in the fit are not entirely independent. In fact, as discussed above, $\sigma^2_{NXS}$ measures the integral of the PSD which, for the long timescale bins, includes some contribution coming from the same high frequencies sampled on short timescales. However this correlation is not expected to be strong, since the PSD has a steep negative slope so that, on each timescale bin, most of the power comes from frequencies close to the minimum sampled value. In any case the correlation would reduce the degrees-of-freedom of our data thus yielding lower probabilities and strengthening our conclusions about which models should be rejected.\\

\subsection{Model fit results}
\label{fit_results}

\begin{table*}
\caption{Fit results with variable Eddington ratio as a function of redshift. The best-fit parameter errors correspond to the 68\% confidence interval.}
\label{fit_varacc_table}
\begin{center}
\begin{tabular}{c|c|c|c|c|c|c|c|c}
\hline\hline
Model & min. $\chi^2_\nu$ & d.o.f. ($\nu$) & Prob ($>\chi^2$) & \multicolumn{5}{c}{best fit $\lambda_{Edd}$}\\
\hline
& & & & $<0.02$ & $0.4\leq z\leq 1.03$ & $1.03< z\leq 1.8$ & $1.8< z\leq 2.75$ & $2.75<z\leq 4$\\
\hline
\multicolumn{9}{c}{Only CDF-S data}\\
\hline
1 & 35.4 & 20 & 0.018 & - & $0.07^{+0.33}_{-0.07}$ & $0.003^{+0.06}_{-0.003}$ & $0.0005^{+0.003}_{-0.000001}$ & $0.0005^{+0.004}_{-0.000001}$\\
2 & 34.5 & 20 & 0.02 & - & $0.17^{+0.21}_{-0.09}$ & $0.09^{+0.08}_{-0.04}$ & $0.04^{+0.04}_{-0.02}$ & $0.03^{+0.05}_{-0.02}$\\
3 & 39.0 & 20 & 0.007 & - & $0.07^{+0.017}_{-0.012}$ & $0.10^{+0.04}_{-0.02}$ & $0.16^{+0.10}_{-0.05}$ & $0.27^{+0.13}_{-0.15}$\\
4 & 30.3 & 20 & 0.07 & - & $0.040^{+0.013}_{-0.009}$ & $0.06^{+0.03}_{-0.02}$ & $0.09^{+0.07}_{-0.03}$ & $0.08^{+0.24}_{-0.05}$\\
\hline
\multicolumn{9}{c}{CDF-S+local AGN \citep{Zhang11} data}\\
\hline
1 & 41.8 & 21 & 0.004 & $ \textit{uncostrained} $ & \multicolumn{4}{c}{\textit{Same as above}}\\
2 & 40.9 & 21 & 0.006 & $ \textit{uncostrained} $ & \multicolumn{4}{c}{\textit{Same as above}}\\
3 & 40.8 & 21 & 0.006 & $0.06^{+0.012}_{-0.009}$ & \multicolumn{4}{c}{\textit{Same as above}}\\
4 & 31.9 & 21 & 0.06 & $0.06^{+0.012}_{-0.009}$ & \multicolumn{4}{c}{\textit{Same as above}}\\
\hline
\end{tabular}
\end{center}
\label{default}
\end{table*}%

Initially we fitted the data allowing a different $\lambda_{Edd}$ for each redshift bin. The best fit models are plotted in Fig.\,~\ref{model_comparison_varacc} and the best-fit results are listed in Table~\ref{fit_varacc_table}, together with the minimum $\chi^2$ and the likelihood of the model. For completeness, we report the best-fit results in the case when we consider the CDF-S data only, and in the case when we add the \citet{Zhang11} data as well. The best-fit $\lambda_{Edd}$ values for the CDFS sources are identical (as each redshift bin is independent from the others), however the addition of Zhang's low redshift data constrain the models better (in terms of the null hypothesis probability). For that reason we discuss the best-fit results when we fit both the CDF-S and the low redshift data. 

From Fig.\,~\ref{model_comparison_varacc} we see that the models become steeper on the shortest timescales, due to the presence of the PSD break, and flatten on the longest one where most of the power comes from the $\nu^{-1}$ part of the PSD. The \citet{Zhang11} lightcurves, on the other hand, are not affected by time dilation due to their redshift $z\simeq 0$, so that the variability is not expected to anti-correlate with luminosity, since the PSD break falls outside the range of probed frequencies..

The best-fit results show that Model 3 is formally rejected at $>99\%$ confidence level. Model 1 provides a statistically acceptable fit, but with an extremely low Eddington ratio, at all redshift bins except the [0.4 - 1.03] bin where it is unconstrained. In fact, the resulting best-fit $\lambda_{Edd}$ values are so low, that even the low luminosity CDF-S AGN should have BH masses larger than $\sim 10^8$ M$_{\odot}$, at all redshifts to explain their observed luminosity. When we force the model to have any value of $\lambda_{Edd}>0.03$ results in the rejection of the model at the $>99\%$ level. We therefore conclude that Models where the break-frequency depends on BH mass only (irrespective of whether the PSD amplitude depends on Eddington ratio or not), are not consistent with the data.  Model 2 is also formally rejected and at the $>99\%$ confidence level. 
We note that Models 1, 2 and 3 all show some tension with the CDF-S data at the lowest luminosities on long timescales (see Fig.\,~\ref{model_comparison_varacc}), as their normalizations are too low.
On the other hand, Model 4 reproduces rather well the overall trends and dependence of variability on luminosity and timescale, for the CDF-S data and its variable PSD normalization allows a better agreement with the observational measurements.

\begin{figure*}
   \centering
   \includegraphics[width=\textwidth]{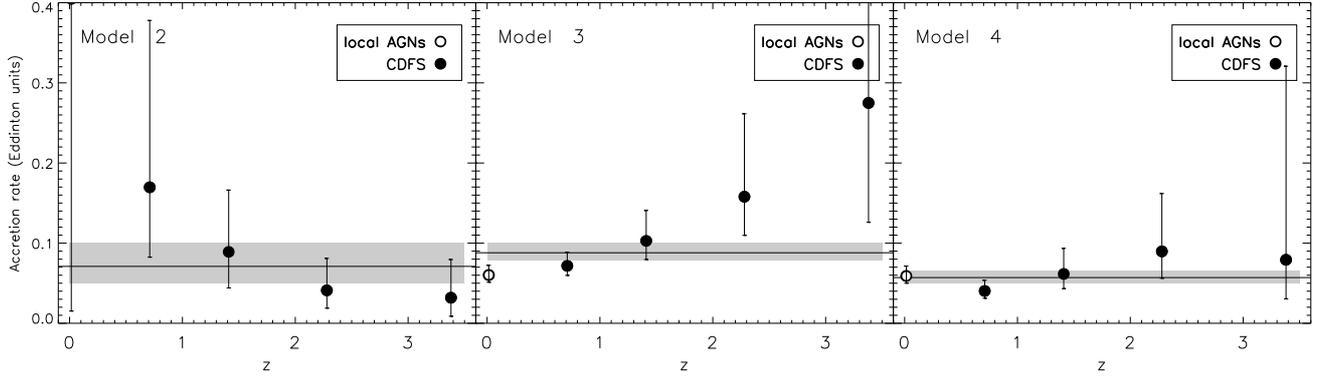}
      \caption{Eddington ratio estimates as a function of redshift from model fitting to the CDF-S and local AGNs variability (Table~\ref{fit_varacc_table}). The solid horizontal lines and shaded areas represent the results of the fit with a constant $\lambda_{Edd}$ reported in Table~\ref{fit_fixacc_Zhang_table}}
         \label{accrate_vs_z}
\end{figure*}

The behaviour of $\lambda_{Edd}$ as a function of redshift is presented in Fig.\,~\ref{accrate_vs_z} (we do not plot the Model 1 best-fit results, as they suggest an accretion rate which is either unconstrained or extremely low). Model 3 predicts an increase of the accretion rate  from $\sim 0.07$ in the local Universe up to almost 0.3 at redshifts higher than 3. However the error of the best-fit parameters are so large, that we cannot claim a significant indication of an increasing accretion rate with increasing $z$. In fact, even in the case of Model 4 (which provides the best-fit to the data)  the best-fit $\lambda_{Edd}$ errors are so large that we cannot argue for a significant variation of the accretion rate with redshift. 

For that reason, we repeated the fits (to both CDF-S and \citealt{Zhang11} data), keeping the accretion rate fixed to a common value at all redshift bins. The best fit results are listed in Table~\ref{fit_fixacc_Zhang_table}. The results are consistent with those presented above. The horizontal solid line in Fig.\,~\ref{accrate_vs_z} indicate the best-fit $\lambda_{Edd}$, which is very similar to the mean of the accretion rate values listed in Table~\ref{fit_varacc_table}. Model 1 fits the data well, but with a very small accretion rate (as before), while Model 3 best-fit is still rejected with a high confidence. This time, Model 2 is marginally accepted(at the 1\% level), but it is still Model 4 which provides again the best fit. In fact, using the $F-$test, we can see that the improvement of the Model 4 best-fit when we let $\lambda_{Edd}$ free is not significant when compared with the best-fit in the case when $\lambda_{Edd}$ is kept constant. 

Summarising, our analysis supports the view that the variability amplitude of the high redshift AGN can be explained if we assume a power spectrum which is identical to the PSD of local AGNs (ie, the variability mechanism is the same in local and high-$z$ AGN). In particular, the variability amplitude of the CDF-S AGN, and its dependence on luminosity, $z$ and time scale, can be explained if the 
PSD break frequency $\nu_b$ depends on both BH mass and $\lambda_{Edd}$, in agreement with \cite{McHardy06}. Most probably, he PSD amplitude also depends on $\lambda_{Edd}$ as proposed by \cite{Ponti12}. The Eddington ratio of the AGN population seems consistent with a constant value, at all redshifts. The quality of our data cannot allow us to detect a dependence of $\lambda_{Edd}$ on $z$.

\begin{table}
\caption{Fit results with constant Eddington ratio using both CDF-S and local AGN \citep{Zhang11} data. }
\label{fit_fixacc_Zhang_table}
\begin{center}
\begin{tabular}{c|c|c|c|c}
\hline\hline
Model & $\chi^2_{min}$ & d.o.f. ($\nu$) & \textit{Prob} ($>\chi^2_{min}$) & best-fit $\lambda_{Edd}$\\
\hline
1 & 44.0 & 25 & 0.011 & $0.0010^{+0.004}_{-0.0008}$\\
2 & 44.0 & 25 & 0.011 & $0.071^{+0.029}_{-0.021}$\\
3 & 51.7 & 25 & 0.0013 & $0.088^{+0.012}_{-0.010}$\\
4 & 34.5 & 25 & 0.10 & $0.057^{+0.008}_{-0.007}$\\
\hline
\end{tabular}
\end{center}
\end{table}%

\section{Discussion and conclusions}
\label{disc_sec}

\subsection{Variability properties of high-redshift AGNs}
In this work we have analysed the lightcurves of AGNs in the CDF-S region, using a dataset spanning $\sim 17$ years. The observing strategy of the CDF-S survey allows us to derive lightcurves with similar sampling for all sources, thus minimising the scatter introduced in timing analysis when different sampling patterns are used for different sources \citep{Allevato13}.
In order to assess the level of variability of our sources we used two different approaches, the first one based on Montecarlo simulations suited to assess whether a source is variable within a certain confidence level, and the second one based on excess variance analysis in order to measure the intrinsic average variability of the AGN population and link it to the physical properties of the AGNs themselves.

We confirm results based on previous studies, that virtually all AGNs are variable, and only the data quality prevents us from detecting variability in faint sources: 90\% of the sources with $>1000$  net counts (e.g. $\sim 20$ counts/epoch) are detected as significantly variable at the $95\%$ confidence level. This result is due in large part to the long timescales probed in the CDF-S dataset, as the likelihood of detecting a source as variable increases with the sampled rest-frame timescale, as expected for sources with a red-noise PSD. In some local AGNs a low-frequency break, has been observed in the PSD, below which the PSD shape becomes flat and the variability power becomes approximately constant. The fact that, on average, the variability of our sources is still increasing on the longest timescales probed by our observations of $\sim 17$ years, constrains the position of this low-frequency break to be, on average over the sampled AGN population, at even longer timescales, in agreement with the results of \cite{Zhang11} and \cite{Middei17}.

One of the most evident clues that AGN variability is related to the physical properties of the central BH is the discovery that variability anti-correlates with intrinsic AGN X-ray luminosity (and possibly also UV/optical luminosity, e.g. \citealt{Collier01,Kelly09,MacLeod10,Simm16}). In X-rays this effect has been observed in samples of nearby AGNs and has been interpreted as the consequence of a larger BH mass in more luminous objects, which would also increase the size of the last stable orbit and thus smear the overall variability produced in the innermost parts of the accretion disk  \citep{Papadakis04}. A clue that such a correlation holds also in higher redshift AGNs came from \textit{ROSAT} observations \citep{Almaini, Manners02}, and has been confirmed by observations of the CDF-S \citep{Paolillo04, Young12, Yang16}, of the Lockman Hole \citep{Papadakis08}, of the COSMOS field \citep{Lanzuisi14} and by serendipitous \textit{XMM-Newton/Swift} samples \citep{Vagnetti11, Vagnetti16}.

In order to study the dependence of variability on the physical properties of the AGNs, we adopted an 'ensemble' analysis approach. Based on the results of the simulation performed by \citet{Allevato13} we focused on a subsample of relatively bright (S/N per bin $>0.8$ or $\gtrsim 350$ counts) sources, in order to avoid introducing statistical biases in our analysis. 
We confirm the presence of an anti-correlation between variability and luminosity, 
but we fail to detect a (statistically significant) high-luminosity upturn suggested in some previous investigations of both the CDF-S sources \citep{Paolillo04} and of other samples \citep{Papadakis08,Vagnetti11}. Our analysis supports the view that high-redshift AGNs have a similar PSD to local sources, where the variability amplitude increases toward longer timescales. Furthermore, while globally the PSD could be represented by a simple power-law model, it seem to be better reproduced by a bending power-law model, similar to many local AGNs where a high-frequency break $\nu_b$ is detected in the PSD. This is possibly the first direct indication that we can extend to high redshift the PSD behaviour observed for local sources.  

Given the complex interplay among variability, luminosity, redshift and timescale, we showed that a proper study of the behaviour of X-ray variability of high-redshift AGNs, its possible evolution and its dependence on the AGN physical parameters (i.e BH mass and accretion rate), must account for all these dependencies simultaneously. To this end we compared the average excess variance, over three orders of magnitude in X-ray luminosity and over 4 different timescales, further dividing the sample in 4 redshift bins, with the predictions from various PSD models. We accounted for both statistical uncertainties and systematic effects due to, e.g., the different timescales and energy ranges probed at each redshift, as well as from the sparse sampling pattern and red-noise leakage, following the recipe of \citet{Allevato13}. To better constrain the variability at low redshift, where the CDF-S survey covers a volume too small to detect a significant number of AGNs (i.e. $z\lesssim 0.02$), we further included in the analysis the sample of local AGN studied by \cite{Zhang11}, whose lightcurves are comparable to those of the CDF-S sources on the longest timescales.  
All models assume a bending power-law where the break frequency $\nu_b$ depends on either the BH mass, or both the BH mass and the Eddington ratio $\lambda_{Edd}$ of the sources \citep{McHardy06}. Additionally we tried models with either a fixed PSD normalization \citep{Papadakis04,Gonzalez-Martin2011}, or a variable one depending on the Eddington ratio $\lambda_{Edd}$ \citep{Ponti12}.

Our results indicate that the variability of high-redshift AGNs is consistent, within the current uncertainties, with PSD represented by a bending power-law where the break frequency $\nu_b$ depends on both the BH mass and the Eddington ratio $\lambda_{Edd}$, in agreement with the results found by \citet{McHardy06} for local sources. Our best fit model suggests that the PSD amplitude should depend on the accretion rate as well, as proposed by \citep{Ponti12}; the tension between models with a fixed PSD amplitude and the CDF-S data on the longest timescales, observed also by \cite{Young12} and by \cite{Yang16}, seem to support this conclusion.
The data rule out the possibility that $\nu_b$ depends only on the BH mass \citep{Gonzalez-Martin2012}, irrespective of whether the PSD normalization is constant or not, unless we adopt implausibly low, average accretion rates. 
We note that understanding the dependence of the PSD on the physical parameters characterizing the AGN population is crucial if we intend to use AGN variability as a cosmological probe, as proposed for instance by \cite{LaFranca14}, since if the measured variability has a dependence on the accretion rate this will introduce an additional scatter in the $M_{BH}-\sigma^2_{NXS}$ relation used to measure cosmological parameters, as well as a systematic redshift-dependent bias if the accretion rate changes with lookback time.

\subsection{Constrains on the BH accretion history}
We used the model fitting results to measure the average accretion history of AGNs. 
To probe a possible change of the Eddington ratio with lookback time we first allowed $\lambda_{Edd}$ to vary, finding that models with fixed PSD normalization show no $\lambda_{Edd}$ dependence on redshift, while models with a variable normalization suggest a possible increase up to $z\sim2-3$. Repeating the fits assuming a constant value for $\lambda_{Edd}$, we obtain that a $\lambda_{Edd}\simeq 0.05-0.10$ is consistent with the data for all the models except one. 

Given the large uncertainties, an increase of the Eddington ratio with lookback time is not ruled out by our data, but in any case the increase cannot be as strong as suggested by previous studies of AGN variability in X-ray selected samples, where it seemed that the high-redshift population at $z>2$ could be dominated by near-Eddington accretors \citep[e.g.][]{Almaini,Manners02,Paolillo04,Papadakis08}. This difference is due in part to the improved statistical approach based on the simulation of \cite{Allevato13}, but also from a new and more physical modelling of AGN variability on different timescales, luminosities and redshift simultaneously, allowed by the extended CDF-S data. This suggests again that extreme care should be used in interpreting variability results based on small, sparse samples without taking into account all sources of bias.  
Arguably, the large uncertainties due to the limited sample size do not yet enable drawing definitive conclusions on the evolution of accretion with redshift but we point out that we are quantitatively constraining $\lambda_{Edd}(z)$ through X-ray variability measurements for the first time. 

The average Eddington ratios derived for the CDF-S sample are in good agreement with estimates in the literature. For instance \cite{Lusso12}, probing AGNs in the XMM-COSMOS survey find $\lambda_{Edd}$ in the range $0.015-0.26$, dependent on both bolometric luminosity and AGN type. Their results are in even better agreement with ours considering that our median $\log(L_{bol})\simeq 45.5$ and for such luminosity the COSMOS sample yields a $\lambda_{Edd}=0.07-0.12$ inclusive of redshift and AGN type dependence. On the other hand \cite{Lusso12} do not find any evidence of $\lambda_{Edd}$ evolution with redshift, although they probe AGNs with $z\lesssim 2.3$ and are thus less sensitive to the higher redshift population than us. \cite{Brightman2013}, exploring the COSMOS and E-CDF-S find Eddington ratios spanning a similar range of Lusso and collaborators, but skewed toward a slightly higher  median $\lambda_{Edd}\sim 0.15$; this is likely a consequence of the somewhat higher median X-ray luminosity probed by their sample compared to this work, since their hard 2-10 keV $\log(L_X)\sim 44.2$ erg s$^{-1}$ while our median value is $\log(L_X)=43.6$ erg s$^{-1}$. Similarly \cite{Suh2015} extend this type of study to a sample including the Lockmann Hole, finding a broad Eddington ratio distribution described by a lognormal distribution peaking at $\lambda_{Edd}\sim 0.25$. Again however this sample extends to higher $\log(L_{bol})\sim 47$ than sampled by our work.
Studies of bright quasars, such as the SDSS-DR7 sample of \citet{Shen2011}, have also usually reported higher average Eddington ratios $\lambda_{Edd}>0.1$  with a sizeable fraction of sources accreting close to the Eddington rate (see, e.g. Figure 4 in  \citealp{Wu2015}); however quasar samples are characterised by bolometric luminosities which are between one and two orders of magnitude greater than all the studies mentioned above, including our own. In the X-ray band luminous quasars have been monitored by \citet{Shemmer14}, finding that their variability is generally larger than expected from their luminosities, based on an extrapolation from lower luminosity sources in the 2 Ms observations of the CDF-S. While this may result from higher average Eddington ratios, it must be noted that unfortunately monitoring observations of such sources in the X-ray band are sparse and heterogeneous, performed through different observatories and in non-dedicated campaigns. This results in large uncertainties that prevented them from drawing definitive conclusions. We further point out that if PSD amplitude depends on $\lambda_{Edd}$, as suggested by our data, the conclusions of  \citet{Shemmer14} (which used fixed PSD amplitude models) may need to be revised in favour of a lower Eddington ratio, although in at least two of their cases the Eddington ratio has been estimated, independently, from H$_\beta$ and the values are quite high. In fact, follow up observations of high-redshift QSO with {\it Chandra} are yielding results consistent with our CDF-S observations (Shemmer et al. 2017, submitted).

 Our variability-based estimates allow us to derive average $\lambda_{Edd}$; however it is clear that AGNs possess an intrinsic Eddington ratio distribution. For instance the work by \citet{Lusso12} derives a lognormal $\lambda_{Edd}$ distribution after correcting for selection effects, while other authors \citep{Aird2012,Bongiorno12,Aird2013} claim that the intrinsic $\lambda_{Edd}$ distribution in galaxies can be represented by a power law function, which is independent of the host galaxy stellar mass, possibly with a soft cutoff around the Eddington limit. In any case the observed $\lambda_{Edd}$ is influenced by selection effects, and tends to display a lognormal distribution peaking at $\lambda_{Edd}\sim 0.01-0.1$.  
A recent revision of the work by Aird and collaborators \citep{Aird17} suggest that the $\lambda_{Edd}$ distribution is more complex than a simple power-law, peaking somewhere in the range $\lambda_{Edd}\sim 0.01-0.1$, although still $\sim$ Eddington-limited, and dependent on host galaxy type (e.g. star-forming vs quiescent). The same study also suggests that the average AGN activity shifts toward higher $\lambda_{Edd}$ with redshift, supporting the tentative trend observed in some of our models. All these results are based on X-ray selected samples of AGNs, but \citet{Vito2016} recently showed that BH accretion in individually X-ray undetected galaxies is negligible compared to the accretion density measured in X-ray sources.
Attempting to constrain the $\lambda_{Edd}$ distribution through variability is difficult due to the large scatter of variability measurements, which is intrinsic to stochastic processes; furthermore  at present our analysis is strongly limited by the available statistics, as we have to divide our data into luminosity and redshift bins. This also limits our ability to explore the possible dependence of  $\lambda_{Edd}$ on other parameters such as the AGN type and host galaxy. Finally, BH spin is the one physical parameter missing from any of the analyses discussed here, so that we are assuming an underlying  spin distribution independent of luminosity, redshift, host-galaxy type etc. Although this is a simplistic approach, unfortunately the study of BH spins is beyond the reach of current facilities, except for a few local AGNs.

We conclude pointing out that the limits of this study are mainly due to the sample size, which prevents us from reducing the statistical uncertainties. Future wide-field/large-effective-area facilities will enable making this method competitive with other tracers, allowing one to probe more effectively the luminosity, redshift and timescale dependence of the intrinsic AGN variability, and also to assess whether the average accretion (and possibly mass) may differ between, e.g., different galaxy populations at each redshift.

\section*{Acknowledgements}
      We thank Fausto Vagnetti and the anonymous referee for the helpful discussions and comments that allowed to improve the manuscript.\\
 This work was supported in part by PRIN-INAF 2014 ``Fornax Cluster Imaging and Spectroscopic Deep Survey''. B.L. acknowledge support from the National Natural Science Foundation of China grant 11673010 and the National Key Program for Science and Technology Research and Development grant 2016YFA0400700. Y.Q.X. and X.C.Z. acknowledge the support from the National Thousand Young Talents program, the 973 Program (2015CB857004), NSFC-11473026, NSFC-11421303, the CAS Strategic Priority Research Program (XDB09000000), the Fundamental Research Funds for the Central Universities, and the CAS Frontier Science Key Research Program (QYZDJ-SSW-SLH006). FEB acknowledges support from CONICYT-Chile (Basal-CATA PFB-06/2007, FONDECYT Regular 1141218), the Ministry of Economy, Development, and Tourism's Millennium Science Initiative through grant IC120009, awarded to The Millennium Institute of Astrophysics, MAS







\bsp	
\label{lastpage}
\end{document}